\documentclass[pra,twocolumn,aps,superscriptaddress,amsmath,amssymb]{revtex4-2}
\usepackage[utf8]{inputenc} 
\usepackage[T1]{fontenc}

\usepackage{lipsum} 
\usepackage{bibentry}
\usepackage[usenames,dvipsnames]{xcolor}
\usepackage{graphicx}
\usepackage[caption=false]{subfig} 
\usepackage{amsmath,amssymb,bm}
\usepackage[version=3]{mhchem}
\usepackage{verbatim}
\usepackage{multirow}
\usepackage{dcolumn}
\usepackage{float}
\usepackage{nicefrac}
\usepackage{siunitx}
\usepackage{booktabs}
\usepackage{chemformula}
\usepackage{wrapfig}
\usepackage{enumitem}  
\usepackage[normalem]{ulem}
\usepackage{transparent}
\usepackage[colorlinks,allcolors=blue,citecolor=blue,urlcolor=blue]{hyperref}
\usepackage{tcolorbox}
\usepackage{ulem}

\definecolor{darkgreen}{rgb}{0.0,0.5,0.0}

\newcommand{\BC}[1]{{\color{black}{#1}}}

\newcommand\tb[1]{\textbf{#1}}
\newcommand\ti[1]{\textit{#1}}

\newcommand\mc[1]{\mathcal{#1}}

\newcommand\beq{\begin{equation}}
\newcommand\eeq{\end{equation}}

\newcommand\im[1]{\textnormal{Im}\left[#1\right]}

\newcommand\dd{\textnormal{d}}

\def\Ha{\mathcal{H}}

\def\x{\tb{r}}
\def\q{\tb{q}}
\def\w{\omega}

\def\v{\tb{v}}

\def\F{\tb{F}}

\def\ne{\delta n_{\rm dr}}
\def\ni{\delta n_{\rm i}}

\def\epsw{\epsilon_{\rm w}}

\def\.{\cdot}

\def\1{^{-1}}
\def\2{^{-2}}
\def\3{^{-3}}

\begin{document}
\title{Fluctuation-induced ionic friction at solid-electrolyte interfaces}

\author{Damien Toquer}
\affiliation{Laboratoire de Physique de l'\'Ecole Normale Sup\'erieure, ENS, Universit\'e PSL, CNRS, Sorbonne Universit\'e, Universit\'e Paris Cit\'e, 24 rue Lhomond, 75005 Paris, France}
\affiliation{Hamburg University of Technology, TUHH, Institute for Physics of Functional Materials (IPFM), Am Schwarzenberg-Campus 3 (E), 21073 Hamburg, Germany}

\author{Baptiste Coquinot}\email{baptiste.coquinot@ist.ac.at}
\affiliation{Laboratoire de Physique de l'\'Ecole Normale Sup\'erieure, ENS, Universit\'e PSL, CNRS, Sorbonne Universit\'e, Universit\'e Paris Cit\'e, 24 rue Lhomond, 75005 Paris, France}
\affiliation{Institute of Science and Technology Austria (ISTA), Am Campus 1, 3400 Klosterneuburg, Austria}

\author{Nikita Kavokine}
\affiliation{The Quantum Plumbing Lab (LNQ), \'Ecole Polytechnique F\'ed\'erale de Lausanne (EPFL), Station 6, CH-1015 Lausanne, Switzerland}

\author{Lyd\'eric Bocquet}\email{lyderic.bocquet@ens.fr}
\affiliation{Laboratoire de Physique de l'\'Ecole Normale Sup\'erieure, ENS, Universit\'e PSL, CNRS, Sorbonne Universit\'e, Universit\'e Paris Cit\'e, 24 rue Lhomond, 75005 Paris, France}

\date{\today}

\begin{abstract}
In this article, we explore how transport at a solid--electrolyte interface is affected by the internal fluctuations of the solid and quantify the role of solid excitations on ionic friction. Combining molecular dynamics simulations with theoretical modeling, we show that the presence of ions enhances the dynamical response of the electrolyte compared to pure water. This leads to an ionic contribution to the fluctuation-induced interfacial friction. We show that collective ionic modes dominate the dissipation, such that ionic friction arises from many-body effects rather than from individual drag forces. Furthermore, high-frequency molecular modes are found to be ion-specific, suggesting possible routes toward spectral separation of ions, with implications for desalination, filtration, and energy conversion.
\end{abstract}

%%%%%%%%%%%%%%%%%%%%%%%%%%%%%%% Main text %%%%%%%%%%%%%%%%%%%%%%%%%%%%%%%
\maketitle

%%%%%%%%%%%%%%%%%%%%%%%%%%%%%%% Introduction %%%%%%%%%%%%%%%%%%%%%%%%%%%%%%%
\section{Introduction}

Ion transport is at the core of many technologies at the water-energy nexus, from filtration, desalination to energy conversion, where permeability, selectivity, and conductivity are tightly coupled~\cite{Tristan2020, Logan2012,  Siria2017}.
Over the past decades, the development of nanofluidic tools has substantially advanced our understanding of water and ion transport in charged nanochannels~\cite{Kavokine2021, Faucher2019, aluru2023fluids}.

In particular, interfacial friction and hydrodynamic slip at solid-liquid interfaces have emerged as key determinants of transport in nanofluidic systems \cite{aluru2023fluids,Kavokine2021}. While interfacial slippage was first rationalized in terms of surface roughness and  hydrophobicity of the surface \cite{Bocquet2010}, recent works have highlighted the role of fluctuation-induced friction as an additional dissipation mechanism. The coupling between charge fluctuations in the liquid and excitation modes of the solid was shown to contribute significantly to fluid friction~\cite{Kavokine2022, Lizee2024, Takeda2025, Kistwal2026, Pryadilin2026}, in particular on some specific materials like graphite, where it is mediated by electronic surface plasmons~\cite{Laitenberger1996, Kavokine2022}. Such effects were shown to account for the counter-intuitive radius dependent slippage in carbon nanotubes \cite{Secchi2016,Kavokine2022}. 

In standard molecular simulations, these effects are typically not accounted for, as electronic excitations of the solid are not treated dynamically within the Born--Oppenheimer approximation. 
One strategy to overcome this limitation is to introduce additional degrees of freedom that mimic the solid's excitation modes~\cite{Herrero2026, Schlaich2022, Bui2023, Sam2025}.
For example, Drude oscillator models attach a fictitious charged particle to each atom via a harmonic spring, thereby introducing a tunable polarization mode with a characteristic frequency~\cite{Lamoureux2003, Misra2017, Bui2023, Sam2025, Kistwal2026}.

Understanding and accurately modeling this fluctuation-induced contribution to friction is therefore essential for predicting and controlling fluid transport at the nanoscale, with direct implications for ionic transport in charged nanochannels. 
Yet, to date, such solid--liquid coupling effects have been investigated primarily for pure water and the effect of dissolved ions remains to be rationalized.

%%%%%%%%%%%%%%%%%%%%%%%%%%%%%%%%%%%%%%%%%%%%%%%%%%%%%%%%%%%%%
\begin{figure*}
\centering	
\includegraphics[width=\textwidth]{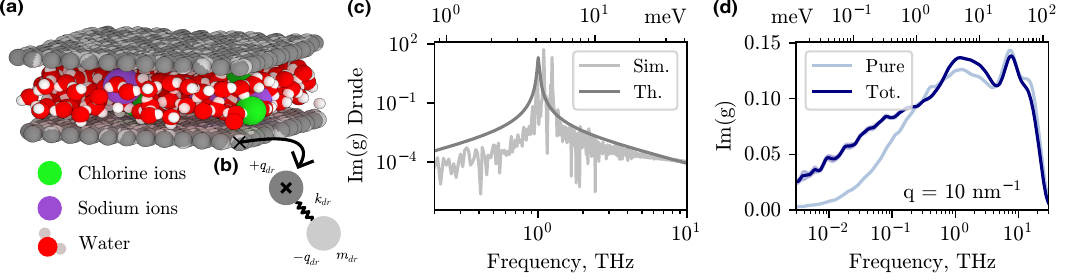}
\caption{\tb{Simulation setup and response functions.} 
\tb{(a)} Slice of the simulation box used in this paper. Chlorine ions are in green, sodium ions are in purple. 
\tb{(b)} Drude polarizable model for the wall. All wall carbon atoms (core atoms) are given a positive charge $Q_{\rm dr}$, and we attach a mobile counter charge $-Q_{\rm dr}$ (Drude atoms), of mass $m_{\rm dr}$ with a spring of stiffness $K_{\rm dr}$.
\tb{(c)} (Light gray) Imaginary part of the surface response function of a single wall of Drude atoms. (Gray) Theoretical prediction (Eq.~\eqref{g_Dr}).
\tb{(d)} Imaginary part of the surface response function (Eq.~\eqref{FDT}) of pure (light blue) and of an electrolyte at 6.4 mol/L (blue) as a function of the frequency. 
 }
\label{fig:fig1}
\end{figure*}
%%%%%%%%%%%%%%%%%%%%%%%%%%%%%%%%%%%%%%%%%%%%%%%%%%%%%%%%%%%%%

Formally, the fluctuation-induced contribution to the fluid-solid friction coefficient,  $\lambda^{\text{FI}}$, can be evaluated from perturbation theory in terms of a spectral overlap between the modes of the solid and the liquid~\cite{Volokitin2006, Kavokine2022}:
\begin{equation}
    \lambda^{\text{FI}} = \frac{k_{\rm B} T}{2\pi^2}\int_0^{\infty}\frac{q^3\dd q\dd\omega}{\omega^2}\frac{\im{g_{\rm s}(q,\omega)}\im{g_\ell(q,\omega)}}{|1-g_{s}(q,\omega)g_{\ell}(q,\omega)|^2},
    \label{lambda_FI}
\end{equation}
Here, $g_{\rm s}$ (resp. $g_\ell$) is the response function of the half-space solid (resp. liquid) to an external electric potential at frequency $\w$ and in-plane wavevector $\q$. 
The liquid response remains challenging to model over the whole momentum and frequency spectrum.
\BC{However,} response functions are directly related to the dynamic structure factors
thanks to the fluctuation-dissipation theorem~\cite{Kavokine2022, Coquinot2023b}. For the liquid (in the $z>0$ half-space), one can write
\begin{equation}
    \im{g_\ell(\q,\omega)} = \frac{e^2}{2\varepsilon_0q}\frac{\omega}{2k_{\rm B} T}\int_{z,z'>0}\! S_\ell(\q,z,z',\omega)e^{-q(z+z')},
        \label{FDT}
\end{equation}
where $z$ is the distance from the interface and 
\beq S_\ell(\tb{r},t;\tb{r'},t')=\langle \delta n_\ell(\tb{r},t)\delta n_\ell(\tb{r'},t')\rangle
 \label{FDT2} \eeq
 is the liquid charge-density dynamical structure factor. A similar expression holds for the solid.

In this article, we investigate the fluctuation-induced friction \BC{with} ions and explore how it is modulated by the solid excitations at an electrolyte-solid interface.
For this purpose, we carry out Molecular Dynamics (MD) simulations of an electrolyte in a 2D nanochannel as shown in Fig.~\ref{fig:fig1}a.
The solid excitations are implemented using Drude oscillators, as depicted in Fig.~\ref{fig:fig1}b, and the resulting fluctuation-induced friction coefficient is computed for various ionic and solid properties.
After detailing the simulation methods, we will quantify fluctuation-induced ion friction and
rationalize this contribution in terms of an overlap between the solid excitations and the fluctuation modes of ions, decomposed into their individual and collective contributions.
Finally, guided by these results, we will develop analytical models for the ionic modes and extend the theoretical formalism of fluctuation-induced friction to electrolytes.

%%%%%%%%%%%%%%%%%%%%%%%%%%%%%%%%%%%%%%%%%%%%%%%%%%%%%%%%%%%%%
\section{Molecular model and solid--liquid surface response}

We consider a 2D nanochannel (Fig.~\ref{fig:fig1}a) of height $h_{\rm slit}=1$~nm, filled with an electrolyte with a given salt concentration $c_{\rm i}$. The strong confinement allows us to focus on ions located in the vicinity of the solid surface, while keeping the numerical complexity tractable.  The water density $\rho_{\rm w} = 21.22~\rm{nm}^{-2}$ was determined from simulations of the slit between reservoirs at $P=1$~atm~\cite{Toquer2025}. Additional information on the system is available in SM~Sec.~3.

The channel walls consist of graphene sheets with atomic density $\rho_{\rm dr}$  per unit surface.
On each carbon atom we place a Drude oscillator consisting of a mobile Drude atom of charge $-Q_{\rm dr}$ and mass $m_{\rm dr}$ attached with a spring of stiffness $K_{\rm dr}$ to a core atom carrying a compensating charge $Q_{\rm dr}$ (see Fig.~\ref{fig:fig1}b). 
Together, this yields a static polarizability:
\begin{equation}
    \alpha_{\rm dr} = \frac{\rho_{\rm dr}Q_{\rm dr}^2}{4\pi\varepsilon_0K_{\rm dr}},
\end{equation}
We use $K_{\rm dr} = 1000$~kcal/mol/\AA$^2$~\cite{Lamoureux2003} and $Q_{\rm dr} = 1.852$~e to reproduce the static electronic polarizability of graphene~\cite{Misra2017}. 
The mass $m_{\rm dr}$ remains a free parameter and is used to tune the harmonic Drude frequency:
\begin{equation}
    \omega_{\rm dr} = \sqrt{\frac{K_{\rm dr}}{m_{\rm dr}}},
\end{equation}
which controls the dynamics of the Drude oscillators.
Indeed, the equation of motion of a single Drude particle reads:
\beq 
m_{\rm dr}\partial_t^2 \x=-K_{\rm dr}\x-2m_{\rm dr}\gamma_{\rm dr}\partial_t \x 
\label{EoM_Drude}\eeq
where $\gamma_{\rm dr}$ is a dissipation term that arises from the thermalization of the Drude atoms.
This parameter is not imposed directly but depends on the thermostatting procedure; in our simulations, we estimate $\gamma_{\rm dr} < \omega_{\rm dr}/100$.

In the continuum limit, Eq.~\eqref{EoM_Drude} becomes an equation of motion for the polarization field and yields the following surface response function for the solid (see~SM~Sec.~2):
\begin{equation}
g_{\rm dr}(\q,\omega)=\frac{ 4\pi \alpha_{\rm dr}\omega_{\rm dr}^2 q }{ \omega_{\rm dr}^2 -\omega^2-2i\gamma_{\rm dr}\omega}e^{-2qd_{\rm dr}} , 
\label{g_Dr}
\end{equation}
A factor of 2 accounts for the transverse mode (see~SM~Sec.~2.3.).
Note furthermore that we introduced a molecular shift $d_{\rm dr}$ to account for the molecular distance between the (arbitrary) plane $z=0$ and the location of the solid-liquid interface; in agreement with previous calculations \cite{Kavokine2022}, we choose  $d_{\rm dr} = 0.13$ nm. 
We have measured the solid surface response $g_{\rm dr}$  in equilibrium simulations using the FDT relationship in Eq.~\eqref{FDT}. In Fig.~\ref{fig:fig1}c, we plot ${\rm Im}[g_{\rm dr}]$ versus the frequency, together with the prediction in Eq.~\eqref{g_Dr}.
The latter is shown to  capture the numerical result for the Drude solid excitation; the small residual discrepancy is attributed to Coulomb interactions between the Drude oscillators.

Turning to the liquid, we compute along the same lines the surface response function on the basis of the charge-charge correlation function and use FDT in Eq.~\eqref{FDT}. Note that equilibrium simulations were performed in the same slit without Drude oscillators, in order to focus on the bare liquid response.
The resulting surface response function, ${\rm Im}[g_{\ell}]$, is plotted in Fig.~\ref{fig:fig1}d, for pure water and for a water-NaCl electrolyte (here with $c_i=6.4$ M). The comparison between pure water and the electrolyte reveals a marked increase of the  response of the electrolyte at sub-THz frequencies in the presence of ions.

%%%%%%%%%%%%%%%%%%%%%%%%%%%%%%%%%%%%%%%%%%%%%%%%%%%%%%%%%%%%%
\begin{figure*}
    \centering	
    \includegraphics[width=\textwidth]{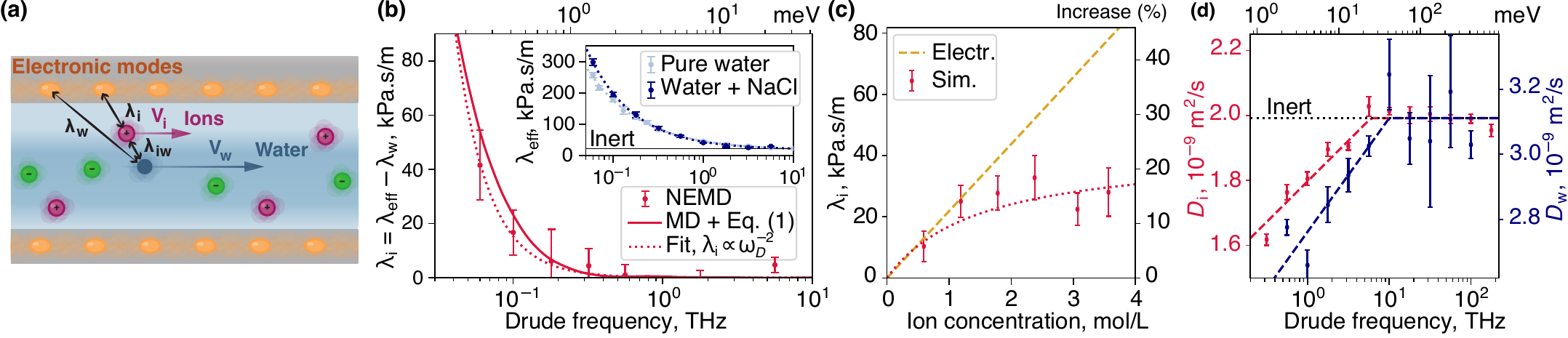}
    \caption{\tb{Ionic friction.}
    \tb{(a)} Sketch of the various contributions to friction for a flowing electrolyte confined between two solids. 
    Water (velocity $v_{\rm w}$) undergoes a friction drag by the solid (friction $\lambda_{\rm w}$), as well as by the ions (friction $\lambda_{\rm iw}$). The ions (velocity $v_{\rm i}$) undergo a friction drag by the solid (friction $\lambda_{\rm i}$) and by  water (friction $\lambda_{\rm iw}$).
     \tb{(b)} Enhancement $\lambda_{\rm i}=\lambda_{\rm eff}-\lambda_{\rm w}$ of solid--liquid friction between an electrolyte at 6.4 M and pure water obtained from non-equilibrium simulations (Red point) \BC{and from equilibrium simulations together with the fluctuation--induced friction formula} Eq.~\eqref{lambda_FI} (Red line). Here, $\lambda_{\rm w}$ was estimated from pure water simulations.
    (Red dotted lines) $\propto\frac{1}{\omega_{\rm D}^2}$ fit.
 Inset: Total solid-liquid friction as a function of the Drude frequency for pure water (light blue) and the electrolyte (blue).  Solid black line is the inert limit $\omega_{\rm dr} \to \infty$. Dotted lines are guides to the eye.
     \tb{(c)} (Red point) Ionic friction for $\omega_{\rm dr} = 0.1$ THz for various ion concentrations obtained from non-equilibrium molecular dynamics simulations. (Red dotted line) Guide to the eye. (Yellow line) Prediction of electronic friction.
\tb{(d)} Diffusion coefficient of ions (average of Na and Cl) and water molecules from MSD simulations. The two diffusion coefficients have been normalized by their limit with an inert solid $\omega_{\rm dr} \to \infty$ (black dashed line).   
}
    \label{fig:fig2}
\end{figure*}
%%%%%%%%%%%%%%%%%%%%%%%%%%%%%%%%%%%%%%%%%%%%%%%%%%%%%%%%%%%%%
%\section{Non-equilibrium simulations and ionic friction}
\section{Transport and ionic friction: (I) single ion contributions}

\subsection{Fluid friction}

We now investigate how the liquid-solid friction coefficient, and in particular its ionic contribution, depends on the solid Drude frequency. To this end, we carry out non-equilibrium Molecular Dynamics (MD) simulations: an external force $f_0$ is applied to the oxygen atoms of all water molecules and the resulting mean water velocity $v_{\rm w}$ is measured. 
This non-equilibrium approach is preferred over equilibrium Green-Kubo methods because strong confinement leads to a rapid decay of force autocorrelation functions, making the extraction of friction coefficients unreliable~\cite{Bocquet1994, Oga2023}.
In all cases we observe a plug flow, which implies that the channel height $h$ is small compared to the slip length $b=\eta/\lambda_{\text{eff}}$, where $\lambda_{\text{eff}}$ is the total solid-liquid friction coefficient and $\eta$ the dynamic viscosity~\cite{Kavokine2021}.
In this situation, the friction coefficient is therefore numerically obtained as
\begin{equation}
   \lambda_{\text{eff}} = \frac{\rho_{\rm w} f_0}{ 2v_{\rm w}},
   \label{def_lambda_eff}
\end{equation}
As anticipated, we find that friction is larger in the presence of an electrolyte than for pure water. This enhancement $\lambda_{\rm eff}-\lambda_{\rm w}$ is shown in Fig.~\ref{fig:fig2}b for a NaCl concentration of 6.4 M, while the corresponding total friction is reported in the inset.
At high Drude frequencies, the difference between the two systems vanishes, and friction is dominated by the atomic corrugation of graphene, which is essentially unaffected by the presence of ions.
At lower frequencies, fluctuation-induced friction becomes dominant. 
In this regime, our results for pure water agree with previous studies~\cite{Bui2023} once confinement effects are taken into account~\cite{Coquinot2023b} (see SM~Sec.~6.2). 
The presence of ions, however, produces a substantial enhancement of friction, concomitant with the stronger low-frequency electric response of the electrolyte (Fig.~\ref{fig:fig1}d). 
Indeed, inserting this enhanced liquid response together with the Drude response of the solid (Fig.~\ref{fig:fig1}c) into the theoretical expression for fluctuation-induced friction, Eq.~\eqref{lambda_FI}, quantitatively reproduces the enhancement observed in Fig.~\ref{fig:fig2}b (solid line). 
We therefore conclude that the ion-induced increase in solid--liquid friction originates from an enhancement of fluctuation-induced friction driven by the amplified low-frequency electric response of the electrolyte.

This enhancement of solid--liquid friction can be understood as follows: the friction of the ions with the solid slows down the ions which in turn slow down the water, as sketched in Fig.~\ref{fig:fig2}a.
This situation can be described by a two-fluid model in which water flows at velocity $v_{\rm w}$ and the ions at velocity $v_{\rm i}$~\cite{Mouterde2018}.
The force balance for water in the steady-state reads: \begin{equation}
    0=\rho_{\rm w} f_0-2\lambda_{\rm w}v_{\rm w}-2\lambda_{\rm iw}(v_{\rm w}-v_{\rm i}),
\end{equation}
where $\lambda_{\rm w}$ is the friction coefficient of water on the solid and $\lambda_{\rm iw}$ characterizes the drag exerted by the ions on the water per unit area.
Meanwhile, the force balance for the ions reads: 
\begin{equation}
    0=-\lambda_{\rm i}v_{\rm i}-\lambda_{\rm iw}(v_{\rm i}-v_{\rm w}),
\end{equation}
where $\lambda_{\rm i}$ is the friction coefficient of the ions on the solid.
Combining these equations yields the effective friction coefficient experienced by water
\beq 
\lambda_{\rm eff} = \lambda_{\rm w}+\frac{\lambda_{\rm iw}\lambda_{\rm i}}{\lambda_{\rm iw}+\lambda_{\rm i}} 
\label{lambdaeff}
\eeq
which identifies with the friction measured numerically via Eq.~\eqref{def_lambda_eff}. 
The first term in \eqref{lambdaeff} corresponds to the direct solid-water friction and the second term represents an ion-mediated contribution, arising from water-ion friction combined with the slowing down of ions by the solid. 

\subsection{Ionic friction}

The total friction experienced by the ions can also be obtained from equilibrium simulations by measuring their mean-square displacement (MSD) at low concentration and for various Drude frequencies.
Indeed, according to the fluctuation--dissipation theorem, the ionic diffusion coefficient satisfies
\begin{equation}
    D_{\rm i} = \frac{k_{\rm B} T}{(\lambda_{\rm i}+\lambda_{\rm iw})/\rho_{\rm i}}.
\end{equation}
where $\rho_{\rm i}$ is the number of ions per unit area.
The results for the diffusion coefficient are reported in Fig.~\ref{fig:fig2}d versus the Drude frequency, showing that diffusion is lowered for low Drude frequencies.

Nevertheless, the total friction experienced by the ions, $\rho_{\rm i} k_{\rm B} T/D_{\rm i}$ is found to be much larger than the effective solid-liquid friction $\lambda_{\rm eff}$ in all cases.
This implies $\lambda_{\rm iw}\gg\lambda_{\rm i}$, corresponding to a much stronger ion--water friction than ion--solid friction. As a result, the velocity difference between water and ions remains small, $v_{\rm w}-v_{\rm i}\ll v_{\rm w}$.
Under this condition, the total friction reduces to the simple addition of the  water and ion contributions, $\lambda_{\rm eff}(c_{\rm i})\simeq \lambda_{\rm w} +\lambda_{\rm i}$.
Thus, assuming $\lambda_{\rm w}$ to be independent of the ion concentration (see SM~Sec.~6.3), the ionic friction exactly corresponds to the solid-liquid friction enhancement presented in Fig.~\ref{fig:fig2}b.
 
For a single ion, the friction arising from its coupling to the electronic modes of a wall, commonly referred to as electronic friction, has already been investigated theoretically~\cite{Persson1995, Tomassone1997, Liebsch1997, Persson2004}. 
In this framework, an ion generates a Coulomb potential that polarizes the solid, with a screening amplitude governed by the surface response function of the solid $g_{\rm s}(\q,\omega)$.
When the ion moves, the induced polarization lags behind, leading to a dissipative force between the ion and its image charge in the solid.
Thus, an ion located at distance $d$ from the Drude wall experiences a force (see~SM~Sec.~4.2):
\beq \label{eq-ef0}
\F_{\rm EF}=-\frac{e^2}{8\pi^2\epsilon_0\epsilon_{\rm eff}}\int\dd^2\q \frac{\q}{q}e^{-2qd}\im{g_{\rm s}^{\rm R}(q,\q\cdot\v_{\rm i})},
\eeq
with $\epsilon_{\rm eff}$ the effective dielectric constant of water~\cite{Bonthuis2012, Becker2025, Fumagalli2018}. 
$\F_{\rm EF}$ is linear in the ion velocity for small velocities.
Using the surface response function for a Drude wall (Eq.~\eqref{g_Dr}) and multiplying by the ion density $c_{\rm i}$, we obtain the electronic friction coefficient:
\beq 
 \lambda_{\rm EF}= \frac{3e^2\alpha_{\rm dr}\gamma_{\rm dr}h}{8\epsilon_0\epsilon_{\rm eff}  \omega_{\rm dr}^2 d_{\rm tot}^4} \times c_{\rm i}
 \label{eq-ef}
\eeq
where $d_{\rm tot}=d+d_{\rm dr}$ is the total distance between the ions and the Drude oscillators.
This model predicts that  the ionic friction scales algebraically with Drude frequency, as 
\beq 
\lambda_{\rm EF}\propto 1/ \omega_{\rm dr}^2,
\eeq
This prediction is in good agreement with the numerical simulations, as shown in Fig.~\ref{fig:fig2}b.

Going beyond, applying this model quantitatively is challenging since the static dielectric function of confined water is a complex function of confinement and furthermore the Drude dissipation $\gamma_{\rm dr}$ is not easily accessible numerically. 
Instead, we adjusted these parameters to fit the low-frequency and low-concentration limit where electronic friction should be valid. 

The electronic friction in Eq.~\eqref{eq-ef} predicts a linear scaling of the ionic friction with the ion concentration, in agreement with the simulation behavior at low salt concentrations, see Fig.~\ref{fig:fig2}c. However, we observe that the measured friction does saturate above $c_{\rm i}\sim1$ M in the simulations, in contradiction with the single-ion model.
This indicates that ionic friction can no longer be described as a sum of independent single-ion contributions at high concentrations, and instead points to the emergence of collective ionic effects.
In particular, many-body ion-ion interactions must be taken into account, consistently with their known importance in ionic conductivity~\cite{Avni2022, Avni2022b, Robin2024, Toquer2025}.

%%%%%%%%%%%%%%%%%%%%%%%%%%%%%%%%%%%%%%%%%%%%%%%%%%%%%%%%%%%%%
%\section{Theoretical framework}
\section{Transport and ionic friction: (II) collective ion contributions}

\subsection{Theoretical framework}

In order to go beyond the model of electronic friction, we extend the formalism of fluctuation-induced friction to include ions~\cite{Kavokine2022}.
We consider a general field theory described by a Hamiltonian
 \beq   \mathcal{H} = \mathcal{H}_0 + \mathcal{H}_{\rm int} \eeq
 where the interaction term describes the electric interactions between the electrolyte and the solid:
  \begin{equation}
    \mathcal{H}_{\rm int}(t) = \int\dd\tb{r}\dd\tb{r}'\delta n_{\ell}(\tb{r},t)V(\tb{r}-\tb{r}')\delta n_{\rm dr}(\tb{r}',t),
\end{equation}
where $\delta n_{\ell}$ and $\delta n_{\rm dr}$ are the charge density fluctuations of the electrolyte (including both water and ions, \BC{$\delta n_{\ell}=\delta n_{\rm w}+\delta n_{\rm i}$}) and the Drude walls respectively, and $V$ is the Coulomb potential.
The charge fluctuations generate a force on the ions, as
\beq  \langle\F_{\rm e\rightarrow i}\rangle(t)=\int \dd\x\dd\x' \,\langle \ni (\x)\nabla_\x V(\x-\x')\ne(\x')\rangle_\Ha. \label{Eq-F}
\eeq
Here, the average is taken with respect to the full Hamiltonian $\Ha$.
The force acting on the ions involves correlations between $\ne$ and $\ni$, whereas the interaction Hamiltonian couples $\ne$ to $n_\ell$.
Consequently, when expanding the force perturbatively in the interaction Hamiltonian, cross-correlations between $\ni$ and $\delta n_\ell$ naturally emerge.
In particular, we find that Eq.~\eqref{Eq-F} gives (see~SM~Sec.~4.4):
\begin{equation}\label{theory_formula}
    \lambda_{\rm i}^{\text{FI}} = \frac{k_{\rm B} T}{2\pi^2}\int_0^{\infty}\frac{q^3\dd q\dd\omega}{\w^2}\frac{\im{g_{\rm dr}(q,\omega)}\im{\mc{A}_{\rm i/\ell}(q,\omega)}}{|1-g_{\rm dr}(q,\omega)g_{\ell}(q,\omega)|^2}
\end{equation}
which generalizes Eq.~\eqref{lambda_FI} to electrolyte systems \cite{Kavokine2022}.
Here, we introduced the spectral function of the cross-correlation function $\mc{A}_{\rm i/\ell}(\q,\w)$ which can be defined through the fluctuation-dissipation theorem as
\begin{equation}\label{eq:crossdef}
\mc{A}_{\rm i/\ell}(\q,\w)=\frac{e^2}{2\epsilon_0q}\frac{i\omega}{2k_{\rm B}T} \langle \ni (\q,\omega) \delta n_\ell(q,\omega)^*\rangle_0,
\end{equation}
while the ionic response describing the ion excitation modes is similarly defined \BC{by Eqs.~\eqref{FDT}-\eqref{FDT2}} as:
\begin{equation}\label{eq:gi}
\BC{\im{g_{\rm i}(\q,\w)}}=\frac{e^2}{2\epsilon_0q}\frac{\omega}{2k_{\rm B}T} \langle \ni (\q,\omega)  \ni(q,\omega)^*\rangle_0.
\end{equation}

%%%%%%%%%%%%%%%%%%%%%%%%%%%%%%%%%%%%%%%%%%%%%%%%%%%%%%%%%%%%%
\begin{figure*}
    \centering	
    \includegraphics[width=\textwidth]{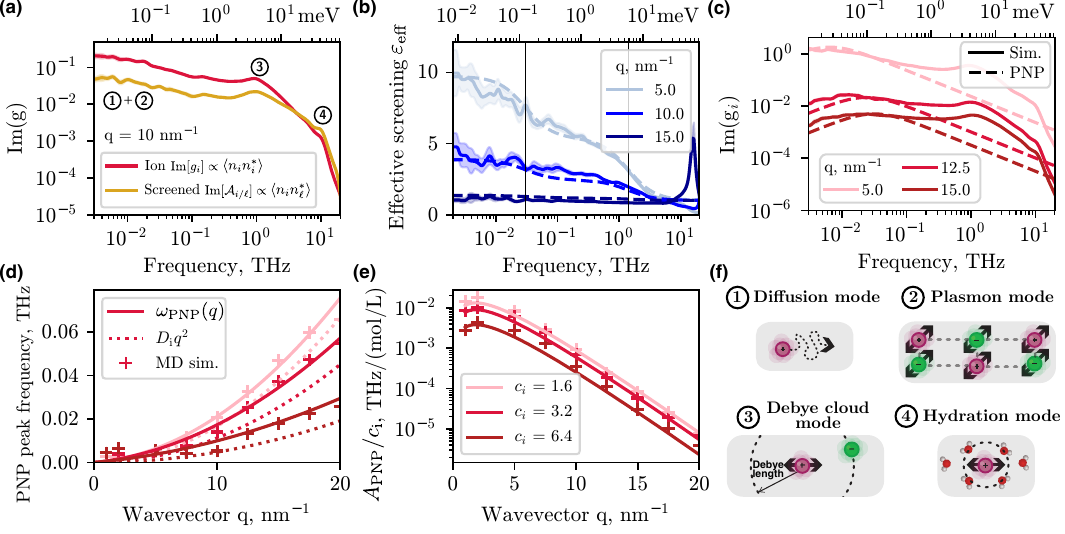}
    \caption{\tb{Ionic excitation modes.}
    \tb{(a)} Imaginary part of the surface response function of ions, $\im{g_{\rm i}(q,\omega)}$ obtained from the correlation function $\langle n_{\rm i}n_{\rm i}^*\rangle$ (red) and of the cross-spectral function, $\im{\mc{A}_{\rm i/\ell}(q,\omega)}$ obtained from the cross-correlation function $\langle n_{\rm i}n_{\ell}^*\rangle$ (yellow), as a function of the frequency at wavevector $q = 10$~nm$^{-1}$. 
  %  (Dashed) Fit of $\im{\mc{A}_{\rm i/\ell}(q,\omega)}$ as $\im{g_{\rm i}(q,\omega)}/\epsilon_{\text{eff}}(q,\omega)$ using the analytic model of $\epsilon_{\text{eff}}(q,\omega)$ (see text).
    \tb{(b)} (Solid lines) Effective permittivity extracted from simulations for various wavenumbers with $c_{\rm i} = 6.4$ mol/L. (Dashed lines) Result of the fit (see text).
    \tb{(c)} Ionic spectrum $\im{g_{\rm i}(q,\omega)}$ (solid lines) versus frequency and for various wavevectors, together with the predictions of the PNP model, Eqs. \eqref{eq:PNP}--\eqref{w_PNP} (dashed lines).
   \tb{(d) and (e)} Respectively the peak frequency and the peak amplitude (divided by the ion concentration) of the low-frequency mode obtained by fitting the surface response functions of ions $g_i$ (dots) and theoretical prediction for the full PNP frequency $\w_{\rm PNP}$ from Eq.~\eqref{w_PNP} (solid lines) and a purely diffusive mode, $\w = D_{\rm i}q^2$ (dashed lines),
    using the diffusion coefficient obtained from MSD simulations. The colors of the lines correspond to the various ion concentrations 1.6 M, 3.2 M and 6.4 M. See SM Table 5 for the diffusion coefficient obtained at these concentrations.
     \tb{(f)} Sketch of the different ionic modes. The labels 1-4 refer to panel \tb{(a)}.
    }
    \label{fig:fig3}
\end{figure*}
%%%%%%%%%%%%%%%%%%%%%%%%%%%%%%%%%%%%%%%%%%%%%%%%%%%%%%%%%%%%%

We calculated these functions in MD simulations of pure water, as well as for an aqueous solution of NaCl ($Z=1$, concentration $c_{\rm i} = 6.4$ M), both with an inert wall, hence without any Drude oscillators. 
Results for the charge correlation functions are given in Fig.~\ref{fig:fig3}a.
As shown on this figure, the two functions $\im{\mc{A}_{\rm i/\ell}}$ and $\im{g_{\rm i}}$ exhibit similar behavior, with the cross response function $\im{\mc{A}_{\rm i/\ell}}$ however lowered below 20 THz. 

The explicit relation between $\im{\mc{A}_{\rm i/\ell}}$ and $\im{g_{\rm i}}$ is theoretically challenging. 
\BC{Still, a simple argument suggests that the former corresponds to the ionic response \emph{after} screening by water.
Indeed, to first order,  the water charge density can be decomposed as $\delta n_{\rm w}(q,\omega) = \delta n_{\rm w}^0(q,\omega) -\chi_{\rm w}(q,\omega) \ni(q,\omega)$, where the first term describes the water charge fluctuations uncorrelated with the ions and the second term is the polarization of water due to the ions, modeled here by a susceptibility $\chi_{\rm w}$.
The cross-correlation, $\im{\mc{A}_{\rm i/\ell}(q,\omega)}$ involves both the water and ion contributions, so that (also see SM~Sec.~4.5):
\begin{eqnarray}
&\im{\mc{A}_{\rm i/\ell}} &\sim \langle \ni  (\delta n_{\rm w}^0 -\chi_{\rm w} \ni)\rangle_0 + \langle \ni \ni^*\rangle_0 \nonumber \\
&& \sim (1-\chi_{\rm w}){\im{g_{\rm i}}}
\end{eqnarray}
Within the random phase approximation (RPA), one has $1-\chi_\w=\epsilon_{\rm w}^{-1}$, with $\epsilon_{\rm w}$ the water dielectric function, and thus we obtain:
\beq
\im{\mc{A}_{\rm i/\ell}(q,\omega)} \approx \frac{\im{g_{\rm i}(q,\omega)}}{\epsilon_{\text{eff}}(q,\omega)}, 
\label{Ansatz}
\eeq
This relationship is approximate, hence the term ``effective'' for the water dielectric function, but it suggests interpreting the ratio of these two functions in terms of the dielectric screening function.
We leave a full derivation of this relation for future investigations.}

%The explicit relation between $\im{\mc{A}_{\rm i/\ell}}$ and $\im{g_{\rm i}}$ is theoretically challenging. Still, a simple argument suggest that the ratio between these two functions is to first order dominated by dielectric screening, \textit{i.e.}
%\beq
%\im{\mc{A}_{\rm i/\ell}(q,\omega)} \approx {\im{g_{\rm i}(q,\omega)}\over \epsilon_{\text{eff}}(q,\omega)}, 
%\label{Ansatz}
%\eeq
%where $\epsilon_{\text{eff}}(q,\omega)$ is the water dielectric function.
%\LB{This relation follows from the fact that the polarization of water due to ions writes as
%$\delta n_{\rm w}^{\rm pol} = -\chi_{\rm w}(q,\omega) \ni(q,\omega)$. Now the cross-correlation, $\im{\mc{A}_{\rm i/\ell}(q,\omega)}$
%involves both the water and ion contributions, so that 
%\begin{eqnarray}
%&\im{\mc{A}_{\rm i/\ell}} &\sim \langle \ni  \delta n_{\rm w}^{\rm pol,*}\rangle_0 + \langle \ni \ni^*\rangle_0 \nonumber \\
%&& \sim (1-\chi_{\rm w}){\im{g_{\rm i}}}
%\end{eqnarray}
%Now, including correlations at the level of the random phase approximation (RPA), one has $1-\chi_\w=\epsilon_{\rm eff}^{-1}$, with $\epsilon_{\rm eff}$ the water dielectric constant, %(for any $(q,\omega)$),
%so that we conclude that $\im{\mc{A}_{\rm i/\ell}(q,\omega)} \approx {\im{g_{\rm i}(q,\omega)}/ \epsilon_{{\rm eff}}(q,\omega)}$.
%This relationship is approximate (hence the term ``effective'' for the water dielectric function) but it interpretes  the ratio of these two functions in terms of the dielectric screening function. We leave a full derivation on this specific point for future investigations.}

Here, we instead compute the (effective) water dielectric constant defined as the ratio 
$\epsilon_{\text{eff}}(q,\omega)\approx {\im{g_{\rm i}(q,\omega)}/ \im{\mc{A}_{\rm i/\ell}(q,\omega)}}$. This function 
is plotted in Fig.~\ref{fig:fig3}b versus frequency for various $q$. Guided by the physical interpretation in terms of the water dielectric
function, we fit $\epsilon_{\text{eff}}$ in terms of the water (Debye) modes, in agreement with previous numerical estimates of the water response~\cite{Kavokine2022}:
\begin{equation}\label{eq:eff_perm_fit}
   \epsilon_{\text{eff}}(q,\omega)-1 = \frac{\epsilon(q,0)-1}{2}\left[\frac{1}{1+\left(\frac{\omega}{\omega_1}\right)^2}+\frac{1}{1+\left(\frac{\omega}{\omega_2}\right)^2}\right],
\end{equation}
Here $f_1 = \omega_1/2\pi = 0.03$ THz coincides with the Debye peak of water; $f_2 = \omega_2/2\pi = 1.4$ THz coincides with the relaxation modes at higher frequency; see details in SM~Sec.~4.6.
Moreover, the low-frequency amplitude $\epsilon_{\text{eff}}(q,0)$ can be fitted empirically by a decaying rational function as
\begin{equation}
    \epsilon_{\text{eff}}(q,0) = \frac{\epsilon_{\text{eff}}(0,0)+c\times q}{1+a\times q+b\times q^2}.
\end{equation}
The values of the numerical parameters are listed in SM~Table~4.
For the system under scrutiny, with confinement $h=1$ nm, we find that $\epsilon_{\text{eff}}(0,0) = 8.87$, which is smaller than in the bulk due to confinement~\cite{Fumagalli2018}. 
As shown in Fig.~\ref{fig:fig3}b, the fit reproduces nicely the frequency and wavevector dependence of the  effective dielectric function $\epsilon_{\text{eff}}(q,\omega)$. 
In this work we did not attempt to directly compute the water dielectric spectrum from the simulations and we leave this for future work.

Now, with this effective dielectric function at hand, we can compute the fluctuation-induced ionic friction directly from the ionic response function, that is from the excitation modes of ions.

%%%%%%%%%%%%%%%%%%%%%%%%%%%%%%% Ionic modes %%%%%%%%%%%%%%%%%%%%%%%%%%%%%%%
\subsection{Low-frequency ionic modes}

\subsubsection{PNP predictions for the ionic collective excitations}

Beyond the individual contribution of ions discussed in the previous section III, we now turn to the collective ionic excitation modes~\cite{Minh2023, Minh2023a}. 
In Fig.~\ref{fig:fig3}a, we observe various characteristic features and peaks in the spectrum of ions for specific frequencies. As we now discuss, these features of the spectrum can be associated with specific modes of the electrolyte and ionic species, as sketched in Fig.~\ref{fig:fig3}f.

At low frequency, the molecular fluctuations of ions can be averaged and continuum hydrodynamics becomes valid~\cite{Bocquet2010}. 
Hence, the ionic charge density $\delta n_{\rm i}$ follows the Poisson-Nernst-Planck (PNP) formalism in which transport is controlled by diffusion and mean-field electrostatic interactions:
\beq \partial_t \delta n_{\rm i} = 2 c_{\rm i}\frac{D_{\rm i}}{k_{\rm B}T}\Delta \phi +D_{\rm i}\Delta\delta n_{\rm i} \label{Eq-PNP} 
\eeq
Monovalent ions have been considered to simplify notations.
Here, $\phi=\phi_{\rm ext}+\epsw ^{-1}\phi_{\rm ind}$ is the electric potential felt by the ions, which consists of the external forcing and the induced electric field generated by the ions. The latter is  screened by the water dielectric function $\epsw$ and is obtained from the Poisson equation, as:
\beq \phi_{\rm ind}(\x,t)=\int\dd\x'\, V(\x-\x') \delta n_{\rm i} (\x',t) \label{Eq-phi_ind}\eeq
where $V(\x)$ is the unscreened Coulomb potential. 
\BC{We introduce the inverse screening length in water $\kappa$} (with relative dielectric constant $\epsw$) and $\kappa_0$ its vacuum counterpart (with relative dielectric constant 1): $\kappa^2=8\pi\ell_B c_{\rm i}$, with $\ell_B=e^2/4\pi\epsilon_0\epsw k_BT$ the Bjerrum length, while
$\kappa_0^2=8\pi\ell_B^0 c_{\rm i}$ with $\ell_B^0=e^2/4\pi\epsilon_0 k_BT$; $c_{\rm i}$ is the bulk ionic concentration.
%\LB{We introduce the screening factors $\kappa$ and $\kappa_0$,  with $\kappa$ the inverse screening length in water (with relative dielectric constant $\epsw$) and $\kappa_0$ its vaccuum counterpart (with relative dielectric constant 1): $\kappa^2=8\pi\ell_B c_{i}$, with $\ell_B=e^2/4\pi\epsilon_0\epsw k_BT$ the Bjerrum length, while
%$\kappa_0^2=8\pi\ell_B^0 c_{i}$ with $\ell_B^0=e^2/4\pi\epsilon_0 k_BT$; $c_{i}$ is the bulk ionic concentration.}
Assuming a quasi-2D ionic layer of surface charge $c_{\rm i}h/2$, Eqs.~\eqref{Eq-PNP}-\eqref{Eq-phi_ind} can be solved in Fourier space to deduce the induced ionic density $\delta n_{\rm i} (\q,\w)$ as a function of the external potential $\phi_{\rm ext}(\q,\w)$. 
The surface response function is then defined as 
\beq 
\phi_{\rm ind}(\q,\w)= -g_{\rm i}(\q,\w)  \phi_{\rm ext}(\q,\w)
\eeq 
and defined at the interface, located at a distance $d$ from the ionic layer.
Accordingly, we obtain, see~SM~Sec.~5.1: 
\begin{equation}\label{eq:PNP}
   g_{\rm i}^{\rm PNP}(\q,\w)= \frac{A_{\rm PNP}(q, c_{\rm i})}{ \omega_{\rm PNP}(q,c_{\rm i})-i\w},
\end{equation}
with an amplitude
\beq
A_{\rm PNP}(q) = \frac{1}{4}D_{\rm i}(c_{\rm i})\kappa_0^2 qh e^{-2qd}
\label{APNP}
\eeq
 and a characteristic frequency
\begin{equation}
    \omega_{\rm PNP}(q,c_{\rm i}) = D_{\rm i}(c_{\rm i})q^2 + \frac{1}{4}D_{\rm i}(c_{\rm i})\kappa^2 qh.
    \label{w_PNP}
\end{equation}
We thus find two ionic modes in Eq.~\eqref{w_PNP} which are sketched in Fig. \ref{fig:fig3}f, as modes 1 and 2 \BC{respectively}. The first term is a diffusive mode of quadratic dispersion, dominant at large wavevectors and low concentrations. The second term is a plasmon mode of linear dispersion coming from inter-ion electrostatic interactions, and dominant at small wavevectors and large concentrations.

In Fig. \ref{fig:fig3}c, we compare the prediction in Eq.~\eqref{eq:PNP} with the spectra obtained from the simulations, focusing here on the low-frequency regime. The simulation results are in fair agreement with the PNP prediction with a broad low-frequency peak varying with the wavenumber $q$. This allows us to extract the characteristic frequency $\omega_{\rm PNP}(q,c_{\rm i})$ (Fig.~3d) and  amplitude $A_{\rm PNP}(q, c_{\rm i})$ (Fig.~3e). These two values for $\omega_{\rm PNP}(q,c_{\rm i})$ and $A_{\rm PNP}(q, c_{\rm i})$  are then compared to the PNP expressions in Eqs.~\eqref{APNP}-\eqref{w_PNP}, showing a good agreement with the theoretical predictions. Note that in these fits we use the value for the diffusion coefficient $D_{\rm i}(c_{\rm i})$ obtained from independent simulations by computing the mean-square displacement of ions (see SM~Table~5).

%%%%%%%%%%%%%%%%%%%%%%%%%%%%%%%%%%%%%%%%%%%%%%%%%%%%%%%%%%%%%
\begin{figure*}
    \centering	
    \includegraphics[width=\textwidth]{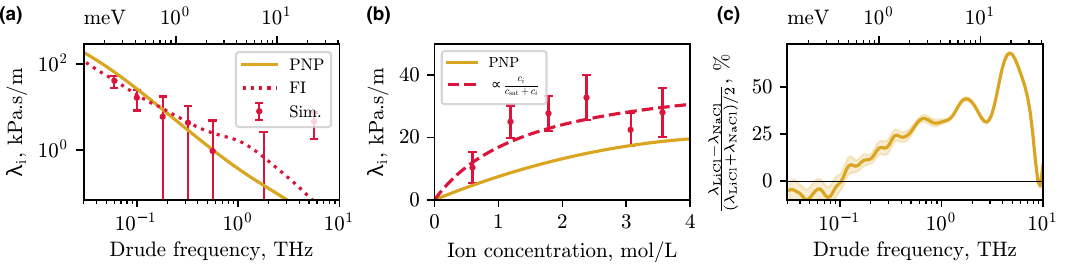}
    \caption{\tb{Theoretical prediction of friction.} 
    \tb{(a)} Comparison between the fluctuation-induced ionic friction Eq.~\eqref{theory_formula} using the simulation spectrum (red dotted line), the prediction using a screened PNP mode Eq.~\eqref{eq:PNP} (yellow solid line) and the direct non-equilibrium simulation (red dot) at various Drude frequencies, with $c_{\rm i} = 6.4$ mol/L. 
    \tb{(b)} Scaling of the prediction of the fluctuation-induced ionic friction with the ionic concentration from our PNP model (yellow solid line, more detail in SM~Sec.~6.1 and 7.3) with the results from NEMD simulations (red dots). The red dotted line is fitted from the theoretical approximation $\lambda(c) = \frac{\delta\lambda c}{c+c_{\rm sat}}$, $\delta\lambda = $ 42 kPa.s/m and $c_{\rm sat} = \frac{1}{2\pi l_{\rm B}d_{\rm tot}h} = 1.49~M$, where $l_{\rm B}$ is the Bjerrum length and $d_{\rm tot} = 0.4$ nm.
    \tb{(c)} Comparison of the result of the fluctuation-induced ionic friction for the NaCl and LiCl spectra, for $c_{\rm i} = 3.2$ M.
    }\label{fig:fig4}
\end{figure*}
%%%%%%%%%%%%%%%%%%%%%%%%%%%%%%%%%%%%%%%%%%%%%%%%%%%%%%%%%%%%%

\subsubsection{PNP excitations and the  fluctuation-induced friction}  

Using these results, we are now in position to compute the effect of the collective excitations on the fluctuation-induced friction.
The PNP expression in Eq.~\eqref{eq:PNP} for the ionic correlation function, together with the ansatz in Eq.~\eqref{Ansatz}, yields a low-frequency prediction for $\im{\mc{A}_{\rm i/\ell}(q,\omega)}$. This is inserted  in Eq.~\eqref{theory_formula} to obtain the fluctuation-induced ionic friction coefficient.

In Fig.~\ref{fig:fig4}a, we compare simulation data (dots) to the PNP contribution (yellow solid line) versus the Drude frequency of the solid. We also plot the total ionic friction obtained from the full ionic spectrum (dotted red line) for comparison.
Notably, in the low-frequency regime, the fluctuation-induced expression quantitatively reproduces both the magnitude of the ionic friction observed in simulations and its scaling as $\lambda_{\rm i}^{\text{FI}}\propto 1/\w_{\rm dr}^{2}$.

This framework also captures the limiting case where the wall frequency greatly exceeds the ionic frequencies: $\omega_{\rm PNP}\ll \omega_{\rm dr}$. In this impurity approximation~\cite{Coquinot2023}, the PNP response function, Eq.~\eqref{eq:PNP}, reduces to a Dirac distribution at zero frequency:
\begin{equation}\label{eq:PNP-limit}
 \frac{1}{\w}  \im{g_{\rm i}^{\rm PNP}(\q,\w)}\approx \frac{A_{\rm PNP}(q, c_{\rm i})}{\omega_{\rm PNP}(q,c_{\rm i})}\pi\delta(\w).
\end{equation}
Further assuming the dilute ion limit where $\omega_{\rm PNP}(q,c_{\rm i})\approx D_{\rm i}q^2$, we recover the electronic friction, Eq.~\eqref{eq-ef0}, by injecting Eq.~\eqref{eq:PNP-limit} into the formula of fluctuation-induced ionic friction, Eq.~\eqref{theory_formula}.

Let us now turn to the effect of ion concentration on the fluctuation-induced friction.
To this end, we estimate the impact of the ionic response on the friction via its integrated spectral weight:
\begin{equation}\label{eq:KK}
\frac{2}{\pi}\int_0^\infty \frac{\im{g_{\rm i}^{\rm PNP}(\q,\w)}}{\w}\dd\w= \frac{A_{\rm PNP}(q, c_{\rm i})}{\omega_{\rm PNP}(q,c_{\rm i})}.
\end{equation}
According to the Kramers--Kr\"onig relation, this quantity also corresponds to the static ionic response
$g_{\rm i}^{\rm PNP}(\q,\omega=0)$.
Since $\kappa^2\propto c_{\rm i}$, we obtain the approximate concentration dependence:
\beq \label{eq:sat} \lambda_{\rm i} \underset{\mathrm{approx}}{\propto}  \frac{2}{\pi}\int_0^\infty \frac{\im{g_{\rm i}^{\rm PNP}(q_0,\w)}}{\w}\dd\w\propto \frac{c_{\rm i}}{c_{\rm sat}+c_{\rm i}} \eeq 
where $q_0\approx1/d_{\rm tot}$ is the typical interaction wavevector set by the ion-wall distance  $d_{\rm tot}$, and $c_{\rm sat}\approx 1/(2\pi\ell_{\rm B}hd_{\rm tot})\approx 1$~M is the saturation ion concentration, with $\ell_{\rm B}$ the Bjerrum length. 

This prediction, shown in Fig.~\ref{fig:fig4}b, reproduces the observed saturation while slightly underestimating the friction due to its simplified screening (see SM~Sec.~6.3 for details).
In the saturation regime, a collective ionic plasmon mode dominates the ionic dynamics. 
The ionic response therefore no longer behaves as the sum of independent ions, hence leading to a saturation of the ionic friction controlled by the collective mode properties.

Thus, our framework allows us to investigate the higher concentration limit where the ionic mobility can no longer be described in terms of individual particles and the notion of excitation modes becomes necessary. 
In particular, these effects could not be observed in Molecular Dynamics simulations with only a few ions as they are many-body effects, just like Coulomb blockade and Wien effects~\cite{Kavokine2019, Robin2021, Toquer2025}.

%%%%%%%%%%%%%%%%%%%%%%%%%%%%%%% Ionic modes %%%%%%%%%%%%%%%%%%%%%%%%%%%%%%%
\subsection{High-frequency modes and ion-specific friction}

While the PNP framework reproduces very well the numerical data for $\w_{\rm dr}< 1$ THz and arbitrary ion concentration, Fig.~\ref{fig:fig4}a shows that it underestimates the ionic friction (dashed line) at larger frequencies. Indeed, in this regime the continuum hydrodynamic picture starts breaking down and molecular effects arise. 
In the spectral perspective developed here, the molecular structure of the electrolyte can be viewed as individual ions trapped in an effective mechanical potential in which they do oscillate, see Fig.~\ref{fig:fig3}f. 

Accordingly, one can model these modes by a sum of harmonic peaks of the form:
\begin{equation}
    g_{\rm i}^{h}(\omega, q) = \frac{A(q)}{\omega_0^2-\omega^2-i\gamma\omega},
\end{equation}
where $\omega_0$ is the harmonic frequency coming from the curvature of the potential, $\gamma$ is the friction coefficient arising from the delay for the mechanical potential to follow the ion and $A(q)$ is the amplitude of the harmonic oscillator. 

A first harmonic mode originates from the interaction of individual ions with the cloud of counter-ions within the Debye length $\lambda_{\rm D}$, which creates an electric potential centered around the ion (see~SM~Sec.~5.3). 
This \ti{Debye cloud mode} is sketched in Fig.~\ref{fig:fig3}f.
In the continuum limit, the electric potential created by the ionic cloud around the central ion of charge $q$ solves the Poisson-Boltzmann equation.
In the Debye--H\"uckel regime of weak electrostatic potentials, assuming spherical symmetry and global electroneutrality, the induced potential reads:
\beq \phi_{\rm dc}(\x) =  \frac{q }{4\pi \epsilon_0\epsw(0) r}\left(e^{-\kappa r}-1\right) .\eeq
We then evaluate the electrostatic energy of the central ion moving within this ionic cloud, modeling the ion as a uniformly charged sphere of radius $a$ to regularize the short-distance singularity:
\beq \mc{E}_{\rm dc}(\x) = \frac{3q}{4\pi a^3}\int_{|\x'|<a}\phi_{\rm dc}(\x-\x') \dd\x' \approx \mc{E}_{\rm dc}(0) + \frac{1}{2}k_{\rm dc}\x^2
\eeq
In the latter expression, $k_{\rm dc}$ denotes the curvature of this energy landscape, from which we deduce the harmonic frequency:
\beq \w_{\rm dc} =\sqrt{\frac{k_{\rm dc}}{m_{\rm i}}}\approx \sqrt{\frac{k_{\rm B}T \ell_{\rm B}}{ 2a\lambda_{\rm D}
^2 m_{\rm i}}} 
\eeq
where $m_{\rm i}$ is the ion mass, and we assumed a small ion size $\kappa a\ll 1$. 
For $c_{\rm i}\approx 6$ M, we find  $\w_{\rm dc} \approx 2$ THz, which indeed corresponds to a harmonic peak in Fig.~\ref{fig:fig3}a. 
The relaxation $\gamma$ comes from the delay for the ionic cloud to follow the ion and is in general frequency-dependent.
Here, the ion oscillates at the harmonic frequency, which is very fast compared to ionic transport, and then the high-frequency limit of $\gamma$ sets the width of the peak. 
However for ionic transport, the low-frequency limit of $\gamma$ would result in a friction coefficient for the ion, ultimately lowering the ionic mobility and increasing the viscosity of the electrolyte~\cite{Avni2022,Avni2022b,Robin2024}.

A second peak comes from the screening by the solvation shell made of oriented water molecules around the ion (see~SM~Sec.~5.4). 
This \ti{hydration mode} is sketched in Fig.~\ref{fig:fig3}f.
The electric potential created by the solvation shell is simply the difference between the potential generated by an unscreened and a screened ion:
\beq \phi_{\rm h}(r)=-\frac{q}{4\pi\epsilon_0 r}\left(1-\frac{1}{\epsw(0)}\right) \eeq
Thus, we can again compute the electrostatic energy of the central ion if it moves in the solvation shell, then its curvature and the resulting harmonic frequency:
\beq  \w_{\rm h} \approx \sqrt{ \frac{k_{\rm B}T\ell_{\rm B} \epsw(0)}{a^3 m_{\rm i}}} \eeq
Here, we used that the dielectric constant of water is large: $\epsw(0) \gg1$.
Altogether,  we obtain $\w_{\rm h} \approx$ 20 THz which matches a corresponding peak in Fig.~\ref{fig:fig3}a. Such a mode has also been reported experimentally~\cite{Schienbein2017, Schwaab2019, Balos2020}. 
Again, the relaxation $\gamma$ comes from the delay for the solvation shell to reorient to follow the ion -- which happens over the Debye frequency -- and is in general frequency-dependent.
At the harmonic frequency, which is much larger than the Debye frequency, the solvation shell is almost still and the relaxation $\gamma$ is small. 
However for ionic transport, the dynamics is much slower than the Debye frequency and the ions experience a much stronger drag force which corresponds to the dielectric friction ~\cite{Bagchi1991, Bagchi1998, Illien2024}.

\section{Some consequences of fluctuation-induced ionic friction }

\subsection{Ion-specific friction and spectral separation}

While the low-frequency PNP modes, reminiscent of hydrodynamics, are mostly universal, the high-frequency modes depend on the molecular structure and are therefore expected to be ion-specific. 
In order to check this prediction, we have compared the ionic friction for two monovalent salts, NaCl and LiCl. 
Accordingly, we performed equilibrium simulations of LiCl at 3.2 M and computed the cross-correlation functions to compute the corresponding fluctuation-induced friction coefficient, $\lambda_{\rm LiCl}$, using Eq.~\eqref{theory_formula}.
The relative difference of the friction predicted for LiCl and NaCl at 3.2 M is shown in Fig.~\ref{fig:fig4}c versus the Drude frequency of the solid.
 As expected, their ionic friction is very close at low Drude frequencies but very different at larger Drude frequencies.

This phenomenon offers the possibility to strongly modify the ionic friction for similar salts by tuning the excitations in the solid. This therefore suggests that these ions could be separated on the basis of ionic friction and specifically via the modulation of its fluctuation-induced contribution. For a proper choice of confining solid material, {\it i.e.} of its spectral properties (here quantified by the Drude frequency), the ionic friction acting on two otherwise similar ions will induce different ionic motion.

This suggests an original methodology for ion separation based on the spectral properties of the confining solid, which has not been explored up to now. One could coin such an approach as ``spectral separation''. Such methodologies remain to be tested experimentally, but applications for ion separation, desalination, or energy conversion would be expected.

\subsection{Friction on physisorbed ions and hydrodynamic slippage}

From a different perspective, the ionic friction may affect electrokinetic transport, in particular in the presence of physisorbed ions at the interface~\cite{Mouterde2018}.
Indeed, it was recently highlighted that surface charging of some material interfaces, such as carbon or boron-nitride surfaces, may occur via ion physisorption instead of the classical chemisorption associated with surface reactivity~\cite{Grosjean2016, Grosjean2019,mangaud2022chemisorbed,wang2025spontaneous}. Physisorption occurs via weak, non-covalent interactions such as van der Waals and electrostatic forces. The physisorbed ions are accordingly bound to the surface, say at a molecular distance $d_{\rm tot}$, but  free to move laterally, parallel to the confining surface~\cite{Grosjean2019,mangaud2022chemisorbed}. The friction experienced by these bound but mobile ions does strongly influence many electrokinetic transport phenomena~\cite{Mouterde2018, mangaud2022chemisorbed}. Accordingly, the spectral properties of the confining solid material would modify the electrokinetic properties. One key property for this purpose is the ionic friction of the physisorbed ions.

\begin{figure}
    \centering	
    \includegraphics[width=0.45\textwidth]{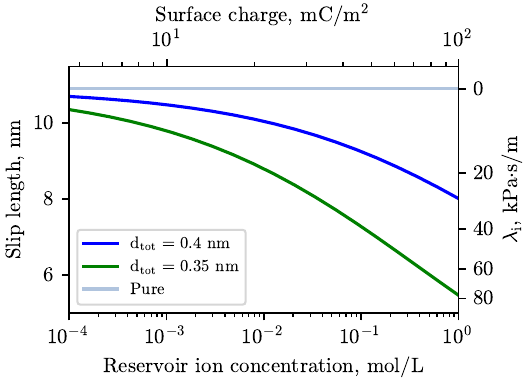}
    \caption{\tb{Impact of the ion-wall distance and of physisorbed ions.} Predicted slip length $b_{\rm eff}$ (left axis) and ionic friction coefficient (right axis) as a function of the reservoir salt concentration $c_s$ for different distances $d_{\rm tot}$ to the walls. (blue) Distance obtained in our simulations. (green) Ions closer to the wall by 0.05~nm. Here, we used $\Sigma=100$ mC/m$^2$ at  $c_{\rm s}=1$ M and $\Sigma \propto c_{\rm s}^{1/3}$. For the bare friction we used $\lambda_{\rm w} = 78$ kPa.s/m extracted from the fluctuation-induced friction of pure water, and not the friction from our NEMD simulation, as it better matches predictions close to an interface (see SM. Sec. 6.2 and Ref. \cite{Bui2023, Coquinot2025}).
    }
    \label{fig:fig5}
\end{figure}

The formalism developed above can be used to describe friction of physisorbed ions at a solid interface. 
While in the previous discussion, we fixed the ion-wall distance via the parameter $d$ entering the surface response functions, it is interesting to vary this distance which accounts for the physisorption strength.
In practice, we evaluate Eq.~\eqref{theory_formula} by varying the ion--wall distance $d_{\rm tot}$ in the ionic surface response function.
As shown in Fig.~\ref{fig:fig5}, the resulting ionic friction is highly sensitive to this distance:
reducing the ion-wall distance  by only $0.05$~nm, we observe a substantial increase in friction by a factor larger than $2.4$.

%%%%%%%%%%%%%%%%%%%%%%%%%%%%%%%%%%%%%%%%%%%%%%%%%%%%%%%%%%%%%
%%%%%%%%%%%%%%%%%%%%%%%%%%%%%%%%%%%%%%%%%%%%%%%%%%%%%%%%%%%%%

As shown by the previous theoretical framework, when individual excitation modes ({\it e.g.}, diffusive modes) dominate the ion--wall coupling, the ionic friction scales with the surface charge, $\lambda_{\rm i} \propto \Sigma$.
At higher surface charge, however, collective plasmon modes become dominant and the ionic friction saturates, as captured by our theoretical model.
Going further, ionic friction will reduce the effective slip length $b_{\text{eff}} = \eta/\lambda_{\rm eff}$, since
\begin{equation}
b_{\text{eff}}(\Sigma) =
\frac{b_0}{1+\frac{1}{\lambda_{\rm w}}\frac{\lambda_{\rm iw}\lambda_{\rm i}}{\lambda_{\rm iw}+\lambda_{\rm i}}}
\approx \frac{b_0}{1+\lambda_{\rm i}(\Sigma)/\lambda_{\rm w}},
\end{equation}
where $b_0=\eta/\lambda_{\rm w}$ is the slip length in the absence of ionic friction. 
In practice, in the presence of surface charge regulation, the surface charge can be experimentally controlled by varying the reservoir salt concentration, with the typical scaling $\Sigma \propto c_{\rm s}^{1/3}$~\cite{Secchi2016b, Green2022, Herrero2024}.
Here, we further show that the characteristic frequency for the solid excitation can impact the ionic friction.

This suggests a route to engineer fluid transport properties via the ionic friction.
For example, assuming a surface charge of $\Sigma=100$ mC/m$^2$ at  $c_{\rm s}=1$ M at the interface of a solid with a characteristic mode at 0.1~THz, the slip length can be tuned by several tens of percent by varying the salt concentration, as shown in Fig.~\ref{fig:fig5}, even though the surface charge remains mobile.

%%%%%%%%%%%%%%%%%%%%%%%%%%%%%%% Conclusion %%%%%%%%%%%%%%%%%%%%%%%%%%%%%%%
\section{Conclusion}

In this article, we have explored the fluctuation-induced friction on ionic solutions close to excitable solids. We have developed a theoretical formalism for the fluctuation-induced friction coefficient and compared its predictions to extensive molecular dynamics simulations of electrolytes confined between excitable solids. 

%To conclude, we showed that adding ions to water enhances the low-frequency electric response of the electrolyte and thereby increases the fluctuation-induced solid--liquid friction. 
%\BC{We find this increase to remain small in strongly confined nanochannels without adsorption, as considered in our simulations. This suggests that, in such systems, fluctuation-induced friction is still dominated by water, with only a minor ionic contribution. However, our simulations provide a general physical picture of ionic friction, allowing our theoretical framework to be extended to systems where mobile ions accumulate near the walls, for instance in the presence of physisorbed surface charges. Adsorption generally brings ions closer to the interface and should therefore strongly enhance the ionic contribution to friction compared to our simulations. In the presence of physisorbed surface charges, ionic friction may become a significant and experimentally tunable contribution to interfacial transport, leading to a substantial reduction of the effective slip length.}
%In particular, nanoscale permeability is not solely an intrinsic property of the solid interface, but can be engineered.
We show that in general the fluctuation-induced ionic friction does not reduce to a sum of independent single-ion friction.
While this impurity picture is recovered in the dilute limit with a solid excitation frequency much larger than the ionic frequencies, it breaks down at high concentrations. 
In this regime, the ionic response becomes collective with a plasmonic mode dominating the ionic response function. 
This collective description rationalizes the saturation of ionic friction observed numerically at high salt concentration and extends the existing theory of fluctuation-induced friction from pure water and isolated ions to concentrated electrolytes.

Our analysis also shows that the ionic spectrum contains both low-frequency collective modes, which are captured by a screened Poisson--Nernst--Planck description, and higher-frequency molecular modes, which depend on the microscopic structure of the electrolyte. 

In particular, the coupling to the ion molecular modes is ion-specific, suggesting that fluctuation-induced friction may acquire a dynamical selectivity for confining solids characterized by high  frequency Drude excitations.
Although this contribution remains subdominant to direct water--solid friction in the systems studied here, it points to the possibility of separating ionic species through their frequency-dependent coupling to the wall in filtration technologies, a phenomenon which we coined spectral separation.

Beyond the fundamental question of solid--liquid friction itself, fluctuation-induced coupling also enables direct energy exchange between the electrolyte and the solid.
Recent work has shown that the coupling between pure water and an electronically active wall can be used for electronic pumping and energy conversion~\cite{Coquinot2023, Lizee2023, Yu2023, Coquinot2024, Coquinot2025, Takeda2025, Herrero2026}.
In parallel, ionic Coulomb-drag effects based on image-charge interactions and electronic friction have been explored in the context of energy conversion~\cite{Rabinowitz2020, Chen2023} and AC transport~\cite{Coquinot2026}.
In this perspective, the direct coupling between ions and solid excitation modes characterized here opens promising routes for energy applications.

%%%%%%%%%%%%%%%%%%%%%%%%%%%%%%%%%%

\section*{Conflict of Interest Statement}

The authors have no conflicts to disclose.

\section*{}
\section*{Data Availability Statement}

The data that supports the findings of this study
are available from the corresponding author upon
reasonable request.

\section*{Acknowledgements}

The authors acknowledge support from ERC project {\it n-AQUA}, grant agreement $101071937$. 
This work was granted access to the HPC resources of MesoPSL financed by the Region Ile de France and the project Equip@Meso (reference ANR-10-EQPX-29-01) of the programme Investissements d'Avenir supervised by the Agence Nationale pour la Recherche. 
B.C. acknowledges support from the CFM Foundation and the NOMIS Foundation. 

\bibliography{bibfile}

%%%%%%%%%%%%%%%%%%%%%%%%%%%%%%% End %%%%%%%%%%%%%%%%%%%%%%%%%%%%%%%
\end{document}

\title{\ti{Supplementary Material for:}\\
Fluctuation-induced ionic friction at solid-electrolyte interfaces}

\author{Damien Toquer, Baptiste Coquinot, Nikita Kavokine, Lyd\'eric Bocquet}

\date{\today}

%%%%%%%%%%%%%%%%%%%%%%%%%%%%%%% Main text %%%%%%%%%%%%%%%%%%%%%%%%%%%%%%%
\maketitle
\tableofcontents
\newpage

%%%%%%%%%%%%%%%%%%%%%%%%%%%%%%%%%%%%%%%%%%%%%%%%%%%%%%%%
\section{Definitions}

%%%%%%%%%%%%%%%%%%%%%%%%%%%%%%%%%%
\subsection{Notations and conventions}

\paragraph{Fourier transform.} We use the Fourier transform:
\beq \tilde f(\q,\w) = \int \dd\x\dd t\, f(\x,t) e^{-i\q\cdot\x+i\w t}, \qquad f(\x,t)=  \int \frac{\dd\q\dd \w}{(2\pi)^{d+1}}\, \tilde f(\q,\w) e^{i\q\cdot\x-i\w t} \eeq

\paragraph{Temperature.} We work at $T=290$ K; therefore, $k_{\rm B}T= 4\e{-21}$ J $=25$ meV.

\paragraph{Ionic density.} We consider ions of charge $\pm Ze$ and density $n_\pm(\x,t)$. We denote $c_{\rm i}=(n_++n_-)/2$ the salt density that we assume to be similar to the bulk salt density. We denote $n_{\rm i} =Zn_+-Zn_-$ the charge density. In average, the charge density vanishes, however, there are fluctuations. As we work in strong confinement, the ionic concentration $c_{\rm i}$ is not perfectly defined.

\paragraph{Ionic lengthscales.}  In the following we denote the Bjerrum length
\beq \ell_{\rm B}(\w)=\frac{e^2}{4\pi \epsilon(\w) k_{\rm B}T}\approx \frac{60\tn{ nm}}{\epsw(\w)}\eeq
 We denote the Debye wavevector
\beq \kappa(\w)=\sqrt{8\pi\ell_{\rm B}(\w)Z^2 c_{\rm i}}\eeq
which is the inverse of the Debye length: $\kappa=\lambda_{\rm D}^{-1}$. 
In bulk water, $\epsw(0)\approx 80$ and we have
\beq \ell_{\rm B}\approx 0.7 \tn{ nm}, \qquad \lambda_{\rm D}\approx \frac{0.3}{Z\sqrt{c_{\rm i}}} \tn{ nm}, \qquad \kappa\approx 3\times Z\sqrt{c_{\rm i}} \tn{ nm}^{-1}, \qquad \tn{ [}c_{\rm i}\tn{ in M]}\eeq

\paragraph{Solvent and dynamics.}  The ions interact with water through a drag: each ion experiences a force $\xi_{\rm i}(\v_{\rm w}-\v_{\rm i})$. The drag coefficient $\xi_{\rm i}$ is related to the ionic mobility and to the ionic diffusion coefficient $D_{\rm i}$ through the fluctuation-dissipation theorem:
\beq \xi_{\rm i}=\frac{k_{\rm B}T}{D_{\rm i}}.\eeq
The corresponding orders of magnitude are:
\begin{center}
$\xi_{\rm i}\sim 10^{-12}$ N$\cdot$s/m and $D_{\rm i}\sim 10^{-9}$ m$^2$/s $\approx 1$ nm$^2$/ns. 	
\end{center}

\paragraph{Coulomb interactions.} The charges interact between themselves with the Hamiltonian:
\beq \Ha_{\rm ab}(t)=\int\dd^3\x\dd^3\x'\, n_{\rm a}(\x,t)V_0(\x-\x')n_{\rm b}(\x',t),\eeq
where $V_0(\x)=e^2/(4\pi\epsilon_0 r)$ is the unscreened Coulomb potential. For a 2D in-plane wavevector, its Fourier transform is $v_q(z)=e^2/(2\epsilon_0 q)e^{-q|z|}$. Its 3D Fourier transform is $v_q^{\rm 3D}=e^2/(\epsilon_0 q^2)$.
In the following, $\phi$ denotes the electric potential, whereas $V$ denotes the Coulomb potential. 
The water and ion velocities are taken into account through a Doppler shift of the frequency argument: $\w\rightarrow \w -\q\cdot\v$.

%%%%%%%%%%% Subsection %%%%%%%%%%%
\subsection{Response functions} 

We consider a solid-liquid interface at $z=0$. 
 In the presence of an external electric potential $\phi\ext(\x,t)$,  the species $\alpha$ induces an electric charge density 
 \beq n_\alpha^{\rm ind}(\x,t)=\int \dd\x'\dd t'\, \chi_\alpha(\x,t,\x',t')\phi\ext(\x',t'), \eeq
 where $\chi_\alpha$ is the susceptibility of the species $\alpha$. 
 Going to Fourier space, we have:
  \beq n_\alpha^{\rm ind}(\q,z,\w)=\int \dd z'\, \chi_\alpha(\q, z, z', \w)\phi\ext(\q, z', \w), \eeq
  where the coordinate $z$ is kept explicit because of the interface at $z=0$. 
  
  This electric charge density induces an (unscreened) electric potential:
  \beq\phi_\alpha(\x,t)=\int\dd\x'\, V_0(\x-\x') n_\alpha(\x',t).\eeq
  Going to Fourier space, by using:
\beq V_0(q, z-z') = \frac{e^2}{2\varepsilon_0q}e^{-q|z-z'|},\eeq we have:
    \beq\phi_\alpha(\q,z,\w)=\frac{e^2}{2\epsilon_0 q}\int\dd z'\,  \chi_\alpha(\q, z, z', \w)\phi\ext(\q, z', \w)e^{-q|z-z'|}.\eeq
    In the following, we are interested in interactions across the solid--liquid interface. Thus, $\phi\ext$ will be generated by a charge density on the other side of the interface and then of the form:
  \beq\phi\ext(\q,z,\w)= \int\dd\x'\, V_0(\x-\x') n_{\rm ext}(\x',t) =  \frac{e^2}{2\epsilon_0 q}\int\dd z'\, n\ext(\q,z',\w) e^{-q|z-z'|}.\eeq
  Hence, the exponential factors take into account distances across the interface. 
  In practice, it is convenient to integrate over the distance from the interface $z=0$ in order to separate properly and consistently each contribution. 
  This leads to the definition of the surface response function \cite{Kavokine2022}:
  \beq g_\alpha(\q,\w) = - \frac{e^2}{2\epsilon_0 q}\int_0^\infty \dd z\dd z'\,  \chi_\alpha(\q, z, z', \w)e^{-q(|z|+|z'|)}. \eeq

%%%%%%%%%%% Subsection %%%%%%%%%%%
\subsection{Elements of Keldysh formalism}

The susceptibilities and surface response functions that we have defined are actually \ti{Retarded} Green's functions. 
In general, it is useful to also define the Advanced and Keldysh versions of the response functions to be able to use the Keldysh perturbation theory \cite{Rammer2007}.
In the following, we only use elements of the computation of \cite{Kavokine2022}, which is carried out in the more general Keldysh formalism, and simply take the classical limit $\hbar\rightarrow 0$. 
Note that in this limit the Keldysh perturbation theory reduces to the standard Martin-Siggia-Rosa action formalism.
To make the applications of perturbation theory transparent, we summarize the relevant elements of the Keldysh formalism in this subsection.

We consider bosonic density operators $n(\tb{r},t)$. We describe the system's dynamics in terms of three types of real-time Green's functions: the Retarded, Advanced and Keldysh Green's functions, defined, for both bosons and fermions, according to 
\beqa\label{definition_Green}
\left\{ \begin{array}{l}
G^{\rm R}(\tb{r},t,\tb{r}',t')= -i \theta(t-t')\langle[ n(\tb{r},t)n(\tb{r}',t') - n(\tb{r}',t')n(\tb{r},t) ]\rangle,\\
G^{\rm A}(\tb{r},t,\tb{r}',t')= i\theta(t'-t)\langle[ n(\tb{r},t)n(\tb{r}',t') - n(\tb{r}',t')n(\tb{r},t) ]\rangle,\\
G^{\rm K}(\tb{r},t,\tb{r}',t')=-i \langle[ n(\tb{r},t)n(\tb{r}',t') + n(\tb{r}',t')n(\tb{r},t) ]\rangle,
\end{array}\right.
\eeqa 
The Retarded and Advanced Green's functions contain information on the system's elementary excitations. The Keldysh Green's function contains information on the quasiparticle distribution. At equilibrium, it satisfies the fluctuation-dissipation theorem: 
\beq\label{DFD}
  G^{\rm K}(\tb{q},\omega)=2 i\,\coth\left(\frac{\hbar\omega}{2k_{\rm B}T}\right)\im{G^{\rm R}(\tb{q},\omega)}
  \eeq
   In the classical limit $\hbar\rightarrow 0$, we obtain the usual fluctuation-dissipation theorem:
   \beq S(\q,\w) = \frac{1}{2i} G^{\rm K}(\tb{q},\omega)=\frac{2k_{\rm B}T}{\w}\im{\chi^{\rm R}(\tb{q},\omega)} \eeq
   where $S$ is the structure factor and $\chi^{\rm R}=G^{\rm R}/\hbar$ is the classical susceptibility. 

For perturbation theory in the Keldysh formalism, we need to consider the matrix form of the Green's function: 
\beqa\label{definition_Green0}
\tb{G}=\left( \begin{array}{cc}
G^{\rm R} & G^{\rm K}\\
0 & G^{\rm A}
\end{array}\right)
\eeqa 
The standard Feynman rules for perturbation theory then apply with this matrix structure, even beyond equilibrium when the fluctuation-dissipation theorem no longer holds.

%%%%%%%%%%% Section %%%%%%%%%%%
\section{Drude oscillators}\label{sec:drude}
%%%%%%%%%%% Subsection %%%%%%%%%%%
\subsection{Surface response function}

We consider a lattice of Drude oscillators at $z=0$ whose electron has a mass $m_{\rm dr}$ and a charge $-Q_{\rm dr}$. Each electron is attached to a fixed ion of charge $Q_{\rm dr}$ through a spring of stiffness $K_{\rm dr}$. We denote $\x_k^0$ the position of the $k$-th ion and $\x_k$ the distance between the $k$-th electron and the $k$-th ion, so that the harmonic force between them is given by
\beq \tb{F}_{\rm ho}^k=-K_{\rm dr}\x_k.\eeq
The oscillator is damped by a drag force
\beq \tb{F}_{\rm th}^k=-2m_{\rm dr}\gamma_{\rm dr}\partial_t \x_k\eeq
In the simulations, this damping is produced by the thermostat; in a physical solid, it would arise from phonons, impurities, and other dissipative processes. 
The only interaction between oscillators is electric. Each oscillator is affected by the local electric potential $\phi$. To compute the response function we apply an external electric potential $\phi_{\rm ext}$. The local electric field is the sum of an external field and the field induced by the oscillators.

The momentum balance of the $k$-th oscillator is
\beq m_{\rm dr}\partial_t^2 \x_k=-K_{\rm dr}\x_k-2m_{\rm dr}\gamma_{\rm dr}\partial_t \x_k+Q_{\rm dr}\nabla\phi(x_k,t). \eeq

We solve this model in the continuum limit. We denote $\tb{u}(\x_k^0,t)=\x_k(t)$ the displacement field and then obtain 
\beq m_{\rm dr}\partial_t^2 \tb{u}=-K_{\rm dr}\tb{u}-2m_{\rm dr}\gamma_{\rm dr}\partial_t \tb{u}+Q_{\rm dr}\nabla\phi(x,t). \eeq
Going to Fourier space we obtain
\beq -m_{\rm dr}\omega^2 \tb{u}=- K_{\rm dr} \tb{u}+2im_{\rm dr}\gamma_{\rm dr}\omega \tb{u} +iQ_{\rm dr} \q\phi(\q,\omega) \eeq
that is 
\beq \tb{u}(\q,\omega)=\frac{iQ_{\rm dr} \q}{ K_{\rm dr}  -m_{\rm dr}\omega^2-2im_{\rm dr}\gamma_{\rm dr}\omega  }\phi(\q,\omega). \eeq

This displacement field $\tb{u}(\x,t)$ generates an electric potential. In the continuum limit, we denote $\rho_{\rm dr}$ the 2D number density of Drude oscillators. At that point, we should distinguish between the longitudinal and transverse modes. For simplicity, we restrict the discussion to the former, the latter being similar. Thus, the induced potential is 
\beq \phi_{\rm ind}(\tb{r},z,t)=\int\rho_{\rm dr}\dd^2\tb{r}'\, Q_{\rm dr}[V(\tb{r}-\tb{r}',z)-V(\tb{r}-\tb{r}'-\tb{u}(\tb{r}',t),z)]\approx \int\rho_{\rm dr}\dd^2\tb{r}'\, Q_{\rm dr}\nabla V(\tb{r}-\tb{r}',z)\cdot \tb{u}(\tb{r}',t)\eeq
where $V$ is the Coulomb potential. Going to Fourier space, we obtain, 
\beq \phi_{\rm ind}(\tb{q},z,\omega)=\frac{\rho_{\rm dr} Q_{\rm dr}}{2\epsilon_0q}e^{-q|z|}i\tb{q} \cdot \tb{u}(\tb{q},\omega)=-\frac{\rho_{\rm dr} Q_{\rm dr}^2}{2\epsilon_0 m_{\rm dr}}\frac{ qe^{-q|z|}}{ \frac{ K_{\rm dr} }{m_{\rm dr}} -\omega^2-2i\gamma_{\rm dr}\omega  }\phi(q,\omega)\eeq
where $n_{\rm dr}(\x,t)=\rho_{\rm dr} Q_{\rm dr}\nabla\cdot\tb{u}(\x,t)$ is the charge density in the material. 
%Denoting $\omega_{\rm p}= \sqrt{K_{\rm dr} /m_{\rm dr}}$ the Drude frequency and $\ell_{\rm p}=\rho Q_{\rm dr}^2/(2\epsilon_0 K_{\rm dr})$ the screening length parametrising the static polarisability $\alpha = \epsilon_0 \ell_{\rm p}/2\pi$, we have 
%\beq \boxed{g_{\rm dr}^{\rm R}(\q,\omega)=\frac{ \omega_{\rm p}^2\ell_{\rm p} q}{ \omega_{\rm p}^2 -\omega^2-2i\gamma\omega}  \label{Drude_response}}\eeq
%At this stage, we have not taken into account the Coulomb interaction between the Drude oscillators. Including them would lead to a dispersion of the Drude frequency that we do not discuss here.

Denoting 
\beq \omega_{\rm dr}= \sqrt{K_{\rm dr} /m_{\rm dr}},\eeq  
\beq \alpha_{\rm dr} = \frac{\rho_{\rm dr}}{4\pi\varepsilon_0}\frac{Q_{\rm dr}^2}{K_{\rm dr}},\eeq the polarizability volume and $d_{\rm dr}$ the distance between the Drude wall and the interface,
we have:
\beq \boxed{g_{\rm dr}^{\rm R}(\q,\omega)=\alpha_{\rm dr}\frac{ 2\pi\omega_{\rm dr}^2 q e^{-2qd_{\rm dr}}}{ \omega_{\rm dr}^2 -\omega^2-2i\gamma_{\rm dr}\omega}  \label{Drude_response}}\eeq
In practice, we add a factor of 2 to account for the presence of a transverse mode in addition to the longitudinal mode described above.
At this stage, we have not taken into account the Coulomb interaction between the Drude oscillators. Including them would lead to a dispersion of the Drude frequency that we do not discuss here.

%%%%%%%%%%% Subsection %%%%%%%%%%%
\subsection{Numerical implementation}\label{sec:drude_num}
To implement the Drude oscillators in our molecular dynamics simulations, we adapt the Drude package in LAMMPS \cite{Thompson2022}. Following \cite{Bui2023}, we arrange the fixed atoms, hereafter referred to as core atoms, on a hexagonal lattice with a core--core distance of $1.42$ \AA\ (yielding $\rho_{\rm dr} = 0.382$ \AA$^{-2}$). We attach a fictitious particle, hereafter referred to as a Drude particle, to each core through a harmonic potential with spring constant $K_{\rm dr} = 1000$ kcal/mol/\AA$^{2}$. The Drude particle has a countercharge $-Q_{\rm dr}$ and a mass $m_{\rm dr}$. We set $Q_{\rm dr} = 1.852$ e to obtain the polarizability volume $\alpha_{\rm dr} = 1.139$ \AA$^{3}$, chosen to reproduce the response of graphite \cite{Misra2017}. Finally, $m_{\rm dr}$ is tuned to vary the Drude frequency $\omega_{\rm dr}$.

To recover the physical classical friction in the $\omega_{\rm dr} \to \infty$ limit, we include no Lennard--Jones interaction involving the Drude particles. We use the core--water Lennard--Jones parameters specified below \cite{Werder2008}. We retain the Coulomb interaction between each core and its Drude particle and apply Thole damping \cite{Thole1981} with a damping parameter of 1.507 to prevent divergences. The Coulomb interaction between the various Drude oscillators leads to a slight difference between the theoretical response in Eq.~\eqref{Drude_response} and the response obtained in simulations. In particular, we observe two distinct peaks near $\omega_{\rm dr}$, each with a slightly different dispersion, as a consequence of the anisotropic interactions within the sheet.

The Drude particles are thermalized using an independent chained Nos\'e--Hoover thermostat with a time constant of $100$ fs and 10 chains. This thermalization determines the Drude relaxation $\gamma_{\rm dr}$, whose value is unknown but small compared with the Drude frequency $\omega_{\rm dr}$.

The interface is chosen to be at $d_{\rm dr}=1.3$ \AA\, from the wall, corresponding to the separating plane between carbon orbitals and water molecules \cite{Kavokine2022}. This choice does not affect the predictions for ionic friction. 

%%%%%%%%%%% Subsection %%%%%%%%%%%
\subsection{Comparison of the simulated and theoretical Drude spectra}
\begin{figure}[H]
    \centering
    \includegraphics[width=0.8\textwidth]{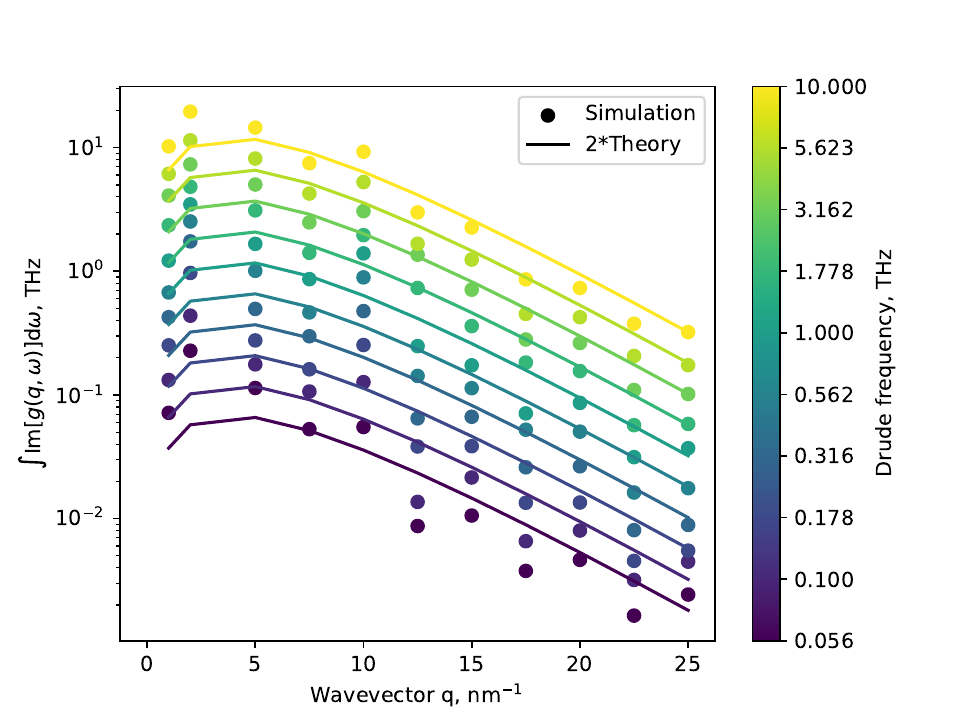}
    \caption{Imaginary part of the surface response function integrated over the frequency axis for the theoretical spectra (solid lines) and simulation data (symbols). The color indicates the Drude frequency. The theoretical spectra are multiplied by a factor 2 to account for the presence of the transverse mode.}
    \label{fig:si_int_drude}
\end{figure}

We simulated Drude walls with various Drude frequencies to test the validity of the theoretical description. In particular, we computed the integrated response
\begin{equation}
    \int_0^{\infty}\operatorname{Im}[g_{\rm dr}(\omega,q;\omega_{\rm dr})]\,\dd\omega
\end{equation}
using Eq.~\eqref{Drude_response}.
The result is shown in Fig.~\ref{fig:si_int_drude}. We find good agreement between the two spectra after multiplying the prediction of Eq.~\eqref{Drude_response} by a factor of 2 to phenomenologically take into account the presence of the transverse mode (in the following, we always include this factor). 
This agreement justifies the use of the theoretical formula to compute the fluctuation-induced friction.
%The difference, in particular at low vector, can be explained by the simple treatment of the interaction between different Drude oscillator in the theoretical model. In particular, we observed that for $q\sim 15$\ nm$^{-1}$ -- which are the mode that matter the more for the fluctuation induced friction -- the Drude spectrum was underestimated by the theory. To correct this, we added a factor 2 in the amplitude (that will be used in all the following formula).

%%%%%%%%%%%%%%%%%%%%%%%%%%%%%%%%%%%%%%%%%%%%%%%%%%%%%
\section{Details of simulations}

\subsection{Simulation setup}

We use LAMMPS \cite{Thompson2022} for all simulations presented in this paper. We consider an electrolyte slab confined between static walls composed of fixed neutral atoms arranged on the same hexagonal lattice as the wall presented in Sec.~\ref{sec:drude_num} of the Supplementary Material. We also use the same force-field parameters for consistency. Both walls are separated by a distance $h_{\rm slit} = 1$ nm. To determine the water density in this geometry, we independently simulated a slit connected to two water reservoirs, with pistons at both ends applying a constant pressure of $P_0 = 1$ atm. We monitored the density in the slit for 500 ps and obtained a final density of $\rho_{\rm w}^{\rm 2D} =$ 0.2122 \AA$^{-2}$.

\begin{figure}[H]
    \centering
    \includegraphics[width=0.8\textwidth]{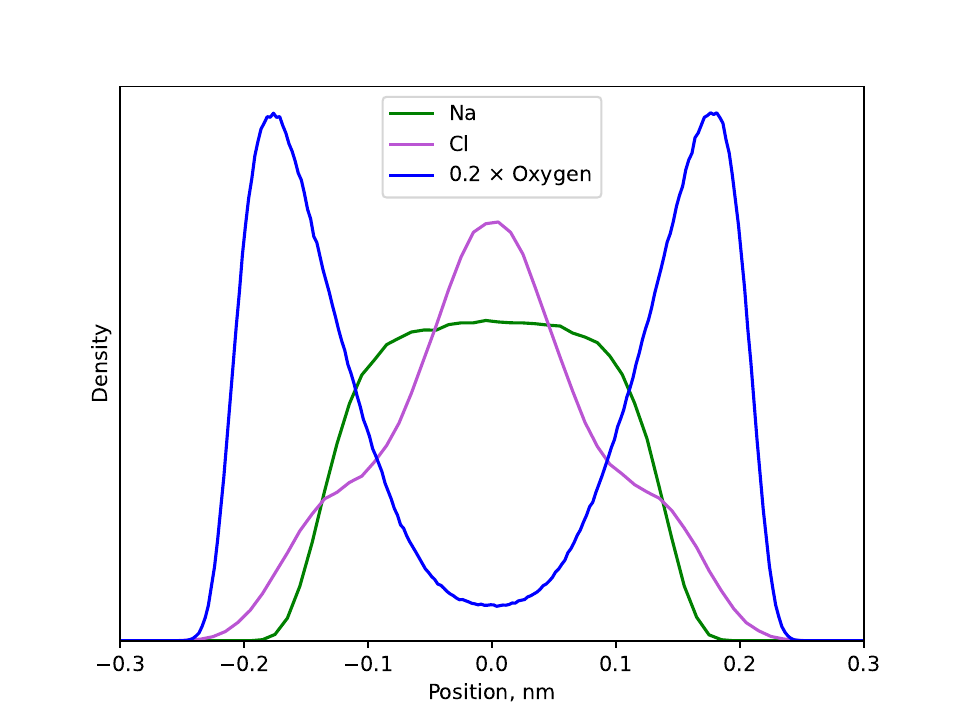}
    \caption{Number-density profiles of oxygen atoms (blue, divided by a factor of 5 for visibility), sodium ions (green), and chloride ions (purple) across the slit.}
    \label{fig:si_density}
\end{figure}

Applying the same procedure in the presence of ions is not very efficient. Instead, we compute a surface ion density $\rho_{\rm i}^{\rm 2D}$ from the ion concentration $c_{\rm i}$:
\begin{equation}
    \rho_{\rm i}^{\rm 2D} = \rho_{\rm w}^{\rm 2D}\frac{c_{\rm i}}{c_{\rm w}},
\end{equation}
where $c_{\rm w} = $ 55.5 mol/L. This definition ensures a consistent number ratio with the bulk, defining an effective height of the slit:
\begin{equation}
    h^{\rm eff} = \frac{\rho_{\rm w}^{\rm 2D}}{\rho_{\rm w}} = 6.35\,\text{\AA}.
\end{equation}
From the surface density, we can compute the number of ions to add to the slit, and we remove the same number of water molecules to compensate. In the main text and in the SM, we denote $h^{\rm eff}$ as $h$, such that we can write $c_{\rm i}^{\rm 2D} = c_{\rm{i}}h$. The density profiles of ions and water across the slit are shown in Fig.~\ref{fig:si_density}.

We use the SPC/E water model~\cite{Berendsen1987}. The force-field parameters are listed in Table~\ref{tab:ff}. Arithmetic (Lorentz--Berthelot) mixing rules are used. The cutoff for the Lennard--Jones and Thole interactions is set to 1 nm.
\begin{table}[H]
    \centering
    \begin{tabular}{ c || c | c | c || c }
        Species & $\varepsilon_{\rm LJ}$ [kcal/mol] & $\sigma_{\rm LJ}$ [\AA] & $q$ [e] & ref\\
        \hline
        \hline
        O & 0.1553 & 3.166 & -0.8476 &  \cite{Berendsen1987}\\
        H & 0 & 0 & 0.4238 &  \cite{Berendsen1987}\\
        \hline
        Na$^+$ & 0.3526 & 2.159 & 1 & \cite{Joung2008}\\
        Li$^+$ & 0.3367 & 1.409 & 1 & \cite{Joung2008}\\
        Cl$^-$ & 0.01279 & 4.830 & -1 &  \cite{Joung2008}\\
        \hline
        C & 0.0567 & 3.214 & 0 & \cite{Werder2008}\\
        Core & 0.0567 & 3.214 & 1.852 & *\\
        Drude & 0 & 0 & -1.852 & *\\
    \end{tabular}
    \caption{\tb{Lennard--Jones and Coulomb parameters.} The C force-field parameters are used for a static wall; for a fluctuating wall, the corresponding core and Drude parameters are used instead.}\label{tab:ff}
\end{table}

For each simulation, the time step is set to 1 fs. All simulations begin with an initial thermalization for a time $T_{\rm th}$ (in the presence of the external force for non-equilibrium simulations). For the electrolyte, we use an independent Nos\'e--Hoover thermostat with a time constant of $100$ fs. The SHAKE algorithm is used for the water molecules.

The system is periodic in all directions, and a 4-nm vacuum gap is added between the periodic slits to avoid interactions between them. The long-range electrostatic interactions are computed using a particle--particle particle--mesh solver with a relative error of $10^{-5}$.

%%%%%%%%%%%%%%%%%%%%%%%%%%%%%%%%%%%%%%%%%%%%%%%%%%%%%

\subsection{Different systems studied}

The different simulation systems are summarized below:

\begin{table}[H]
    \centering
    \begin{tabular}{ c || c | c | c || c | c | c }
        System & Type & Usage & Wall & $L_x$ (\AA) & $L_y$ (\AA) & $c_{\rm i}$ (M)\\
        \hline
        \hline
        I & Equi. & Spectrum & Inert & 63.96 & 63.9 & 0, 1.6, 3.2, 6.4\\
        II.(a) & NEMD & Slip length & Drude, $\omega$ variable & 147.6 & 149.1 & 0, 3.2, 6.4\\
        II.(b) & NEMD & Slip length & Drude, $\omega$=0.1 THz  & 147.6 & 149.1 & Variable\\
        III & Equi. & Diffusion & Drude, $\omega$ variable  & 147.6 & 149.1 & 0, 0.024\\
    \end{tabular}
    \caption{\tb{General information on the systems studied.} ``Equi.'' denotes equilibrium simulations without external forcing, whereas NEMD denotes simulations driven by an external force applied to each oxygen atom.}\label{tab:syst}
\end{table}

The thermalization times, production times, and numbers of independent simulation runs are listed below:

\begin{table}[H]
    \centering
    \begin{tabular}{ c || c | c | c }
        System & $T_{\rm th}$ (ns) & $T_{\rm sim}$ (ns) & $N_{\rm sim}$ \\
        \hline
        \hline
        I & 0.2 & 10 & 1\\
        II.(a) & 0.2 & 0.8 & 3\\
        II.(b) & 0.02 & 0.6 & $\simeq$8\\
        III & 0.2 & 1 & $\simeq$3\\
    \end{tabular}
    \caption{\tb{Additional details on the systems studied.}}\label{tab:stat}
\end{table}

%%%%%%%%%%%%%%%%%%%%%%%%%%%%%%%%%%%%%%%%%%%%%%%%%%%%%

\subsection{Spectrum and surface response function}
All spectral calculations use trajectories from system I (see Tab.~\ref{tab:syst}). The trajectories are sampled every 10 fs, thereby satisfying the Nyquist--Shannon sampling criterion. Each trajectory is divided into 10 subtrajectories, and the uncertainty is estimated from the standard error among the resulting spectra.
For any species $\alpha$, at each time $t$, we can compute a surface charge density according to:
\begin{equation}
    n_{\alpha}(\tb{q},z,t) = \sum_{i\in\alpha}q_{i}e^{i\tb{q}\cdot\tb{r}_i(t)}e^{-q|z_i(t)-z_0|},
\end{equation}
where $\tb{q}$ is the 2D in-plane wavevector, $q$ is its norm, $\tb{r}_i$ is the position of any charge $i$ in the species $\alpha$ at $t$, and $z_0$ is chosen to be at $1.3\ \textrm{\AA}$ from the wall. The wavevectors are constrained by the size of the box:
\begin{equation}
    \tb{q}_{m,n} = \left(m\frac{2\pi}{L_x},n\frac{2\pi}{L_y}\right),
\end{equation}
where $L_x$ and $L_y$ are the dimensions of the box, and $m$, $n$ are integers. In practice, for various values of $q$, we average over multiple pairs $(m,n)$ that satisfy:
\begin{equation}
    |q_{m,n}-q|<0.02\ \textrm{\AA}^{-1}.
\end{equation}

We then define the fluctuation of the charge density:
\begin{equation}
    \delta n_{\alpha} = n_{\alpha}-\langle n_{\alpha}\rangle,
\end{equation}
where $\langle\cdot\rangle$ denotes an ensemble average, or a time average (by ergodicity).

For two species $\alpha$ and $\beta$, we can define a cross-correlation function:
\begin{equation}
    S_{\alpha,\beta}(\tb{q},t) = \left\langle\delta n_{\alpha}(\tb{q},t)\delta n^*_{\beta}(\tb{q},0)\right\rangle,
\end{equation}
Using the Wiener--Khinchin theorem,
\begin{equation}
    S_{\alpha,\beta}(\tb{q},\omega) = \frac{\delta n_{\alpha}(\tb{q},\omega)\delta n^*_{\beta}(\tb{q},\omega)}{T_s},
\end{equation}
where $T_s$ is the total simulation time. If $\alpha=\beta$, $S$ is real valued. Otherwise, we take the real part of the quantity.

Finally, we define the cross surface response function:
\begin{equation}
    \im{g_{\alpha,\beta}(q,\omega)} = \frac{e^2}{2\varepsilon_0q}\frac{\omega}{2k_{\rm B}T}S_{\alpha,\beta}(\tb{q},\omega).
\end{equation}

The spectra are smoothed using a Gaussian filter with a standard deviation $\delta\omega = \omega/10$. We checked that this choice does not lead to unwanted artifacts. 

%%%%%%%%%%%%%%%%%%%%%%%%%%%%%%%%%%%%%%%%%%%%%%%%%%%%%

\subsection{Friction coefficient from non-equilibrium simulations}

In order to compute the friction coefficient, we perform non-equilibrium molecular dynamics simulations with various driving forces $f_0$. We apply the force only to the oxygen atom of each water molecule. Systems II.(a) and II.(b) are used (see Tab.~\ref{tab:syst}).

\begin{figure}[H]
    \centering
    \includegraphics[width=0.8\textwidth]{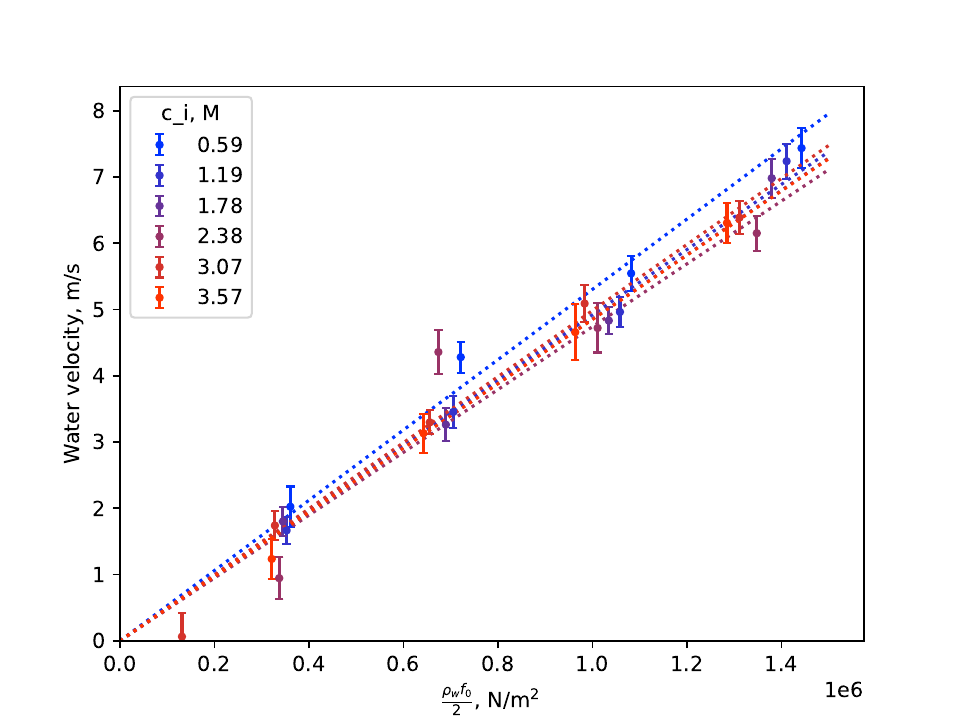}
    \caption{Average steady-state water velocity for various external forces. The dotted lines are the corresponding linear fits used to obtain $\lambda_{\rm eff}$.}
    \label{fig:si_linearity}
\end{figure}

For the simulation performed on system II.(b), we use $\omega_D = 0.1$~THz and vary the ionic concentration between 1 and 4 M. For each simulation, to check the linearity of the response, the water steady-state velocity is computed for $f_0 \in \{0.0005, 0.001, 0.0015, 0.002\}$~kcal/mol/$\textrm{\AA}^{-1}$, starting from two independent initial states to improve the statistics. The standard error on each trajectory is estimated using the pyblock Python library \cite{githubGitHubJsspencerpyblock} and from the standard error between the independent simulations. The results are shown in Fig.~\ref{fig:si_linearity}.

For the simulation performed on system II.(a), because the friction varies strongly with the Drude frequency, we increase the force with the friction to retain sufficient statistical accuracy while remaining in the linear regime. We check that linearity is maintained in all simulations as long as $v\lesssim 12$ m/s and use this as a criterion. In practice, the force is varied between $0.0003$ kcal/mol/$\textrm{\AA}^{-1}$ at the lowest frequency and $0.003$ kcal/mol/$\textrm{\AA}^{-1}$ at the highest. The standard error on each trajectory is estimated using the pyblock Python library \cite{githubGitHubJsspencerpyblock} and from the standard error between the independent simulations.

%%%%%%%%%%%%%%%%%%%%%%%%%%%%%%%%%%%%%%%%%%%%%%%%%%%%%%%%

\subsection{Diffusion coefficient and conductivity}\label{sec:si_diff}

We also performed equilibrium simulations of water and ions in the presence of Drude particles in order to compute equilibrium quantities and the conductivity through the Green--Kubo formula. This was done using trajectories from system III (see Tab.~\ref{tab:syst}). We also computed the diffusion coefficient without Drude particles at higher ion concentrations using trajectories from system I.

To compute the diffusion coefficient, we considered only the $x$ and $y$ components of the atomic positions and computed the mean-square displacement:
\begin{equation}
    \operatorname{MSD}(\tau) = \left\langle\left|\tb{r}_i(t_0+\tau)-\tb{r}_i(t_0)\right|^2\right\rangle_{i,t_0} \sim 4D\tau,
\end{equation}
averaged over the initial times and the particles.

For the ions from systems III and I, we used the tidynamics library \cite{Buyl2018}. For system III, the diffusion coefficient is extracted by fitting the MSD between $\tau$ = 0.3 ps and $\tau = 10$~ps to a linear function. For system I, as we can access longer simulation times, we fitted the MSD between $\tau$ = 10 ps and $\tau$ = 1 ns. For system III, the error is obtained from the standard error between independent trajectories and different ions (assumed independent). 

For the water from system III, we directly computed the MSD, and the diffusion coefficient is extracted by fitting the MSD between $\tau$ = 0.2 ps and $\tau = 0.8$~ps to a linear function.

We then computed the molar conductivity using the Green-Kubo formula:
\begin{equation}
    \Lambda_\alpha = \lim_{t\to\infty}\frac{V}{c_{\rm i}\,k_{\rm B}T}\int_0^{t}\dd s\langle j_\alpha(0)j_\alpha(s)\rangle_{E=0},
\end{equation}
where $j_{\alpha}$ is the electric current density created by species $\alpha$:
\begin{equation}
    j_{\alpha} = \frac{1}{V}\sum_{i\in\alpha}{q_{\alpha}}v_i.
\end{equation}
The running integral of $\Lambda_{\alpha}$ reaches a plateau between 3 and 10 ps. The conductivity is defined as the average of $\Lambda$ over this plateau.

\begin{figure}[H]
    \centering
    \includegraphics[width=\textwidth]{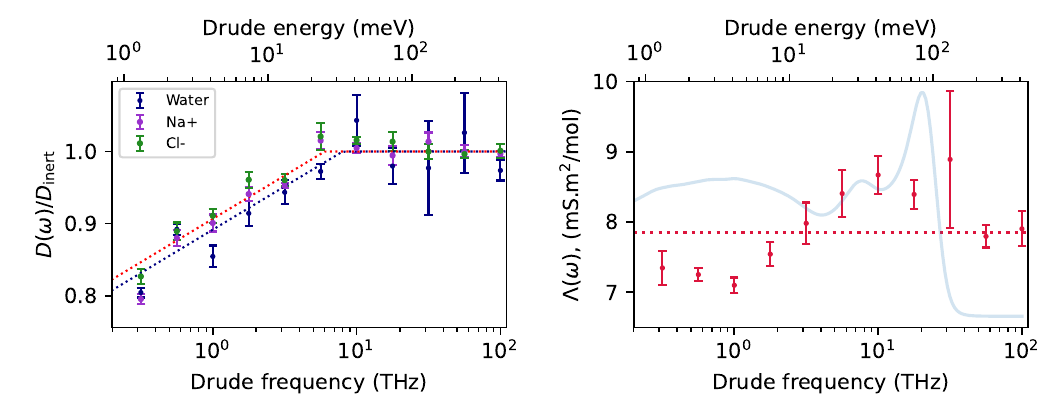}
    \caption{\tb{Diffusion and conductivity.} Left: diffusion coefficients of the cations (purple), anions (green), and water molecules (blue) as functions of the Drude frequency. Dotted lines are guides to the eye. $D_{\rm inert}$ is the average value for $\omega_D>1$~THz. For $\mathrm{Cl}^-$, $\mathrm{Na}^+$, and water, respectively, $D_{\rm inert} = 2.16\times 10^{-9}$, $1.82\times 10^{-9}$, and $3.11 \times 10^{-9}$ m$^2$/s. Right: molar conductivity as a function of the Drude frequency.}
    \label{fig:si_diff_cond}
\end{figure}

The results are shown in Fig.~\ref{fig:si_diff_cond}. As discussed in the main text, diffusion is reduced by slow Drude oscillators. We observe a difference in behavior between the ionic self-diffusion and the conductivity. Comparison with the pure-water spectrum suggests that the modulation of the conductivity at high frequency, $f\sim 1$ THz, is a consequence of water--wall coupling. Near the Debye peak of water, we observe a decay similar to that of the self-diffusion coefficient, which appears to saturate at lower frequencies. This could motivate a more thorough investigation, which is beyond the scope of this paper. Nevertheless, these results show that the wall excitation modes and their coupling to the electrolyte may play a key role in ionic transport through nanochannels and should be taken into account to optimize conductivity.

%%%%%%%%%%%%%%%%%%%%%%%%%%%%%%%%%%%%%%%%%%%%%%%%%%%%%%%%
\section{Perturbation theory and screening}

%%%%%%%%%%%%%%%%%%%%%%%%%%%%%%%%%%%%%%%%%%%%%%%%%%%%%%%%
\subsection{Model}

We consider a system made of a solid in $z<0$ with an electronic density $\ne$ and a Hamiltonian $\Ha_{\rm s}$ and an electrolyte in $z>0$ with a water charge density $\nw$ and an ionic density $\ni$. The interactions inside the electrolyte are described by a Hamiltonian $\Ha_\ell$, which includes the solvent's internal structure, the ion--ion Coulomb interactions and the ion--solvent interactions, namely contact forces and solvation effects. The solid and the liquid interact through two Coulomb interactions, namely:
\beq \Ha_{\rm w/s}=\int \dd\x\dd\x' \, \nw(\x)V(\x-\x')\ne(\x')\eeq
\beq \Ha_{\rm i/s}=\int \dd\x\dd\x' \, \ni (\x)V(\x-\x')\ne(\x')\eeq
 The total Hamiltonian is then
\beq \Ha=\Ha_{\rm s}+\Ha_\ell+\Ha_{\rm w/s}+\Ha_{\rm i/s}.\eeq 

The force applied by the solid on the ions is
\beq 
\boxed{
 \langle\F_{\rm e\rightarrow i}\rangle(t)=\int \dd\x\dd\x' \,\langle \ni (\x)\grad_\x V(\x-\x')\ne(\x')\rangle_\Ha }
 \label{force_general}
\eeq
where the average is taken over the Hamiltonian $\Ha$. 

%%%%%%%%%%%%%%%%%%%%%%%%%%%%%%%%%%%%%%%%%%%%%%%%%%%%%%%%
\subsection{Electronic friction}

For a single ion, the resulting force is called electronic friction. For an ion moving with velocity $\v$ at a distance $d$ from the wall, Eq.~\eqref{force_general} reads:
\beq \F=-\frac{e^2}{8\pi^2\epsilon_0\epsw}\int\dd\q \frac{\q}{q}e^{-2qd}\im{g_{\rm e}^{\rm R}(q,\q\cdot\v)}\eeq
where we have taken into account the screening of the Coulomb interaction by water. 
At first order in the velocity, we then obtain the electronic friction drag coefficient: 
\beq \xi_{\rm EF}=\frac{e^2}{8\pi\epsilon_0\epsw}\int\dd q q^2e^{-2qd}\partial_\omega\im{g_{\rm e}^{\rm R}(q,\omega=0)}\eeq

For a surface ionic density $2c_{\rm i}h$, the total force reads:
\beq \boxed{ \lambda_{\rm EF}=\frac{e^2}{4\pi\epsilon_0\epsw}c_{\rm i}h\int\dd q q^2e^{-2qd}\partial_\omega\im{g_{\rm e}^{\rm R}(q,\omega=0)} }\eeq
For Drude oscillators, we can use Eq. \eqref{Drude_response}.
Then,
\beq \partial_\omega\im{g_{\rm e}^{\rm R}(q,\omega=0)} =\alpha_{\rm dr}\frac{ 4\pi \gamma_{\rm dr} }{ \omega_{\rm dr}^2}q e^{-2qd_{\rm dr}}  \eeq
and, denoting  $d_{\rm tot}=d+d_{\rm dr}$ the total distance between the ion and the Drude oscillators, the electronic friction reads:
\beq  \lambda_{\rm EF}=\frac{e^2\alpha_{\rm dr} \gamma_{\rm dr} }{\epsilon_0\epsw  \omega_{\rm dr}^2}c_{\rm i}h\int\dd q q^3e^{-2qd_{\rm tot}} = \frac{3e^2\alpha_{\rm dr}\gamma_{\rm dr}}{8\epsilon_0\epsw  \omega_{\rm dr}^2 d_{\rm tot}^4}c_{\rm i}h
\eeq

%%%%%%%%%%%%%%%%%%%%%%%%%%%%%%%%%%%%%%%%%%%%%%%%%%%%%%%%
\subsection{Perturbation theory with screening by hand}

We now consider charge fluctuations and compute the force on ions perturbatively. We take: 
\beq \Ha_0=\Ha_{\rm s}+\Ha_\ell+\Ha_{\rm w/s} \quad \tn{ and }\quad \Ha_{\rm int}=\Ha_{\rm i/s}\eeq
In this version, the solid still interacts with the liquid and then indirectly with the ions under $\Ha_0$. 

In linear response theory, the force on the ions becomes:
\beq \langle\F_{\rm e\rightarrow i}\rangle(t)=\int \dd\x\dd\x'\dd t' \,\grad_\x\langle[\ni (\x)V(\x-\x')\ne(\x'),\Ha_{\rm int}(t')]\rangle_0\eeq
where the average is now taken over the Hamiltonian $\Ha_0$.
Thus, 
\beq \langle\F_{\rm e\rightarrow i}\rangle(t)=\int \dd\x\dd\x'\dd\bar\x\dd\bar\x'\dd t'\grad_\x V(\x-\x')V(\bar\x-\bar\x')\langle [\ni (\x)\ne(\x'),\ni (\bar\x)\ne(\bar\x')]\rangle_0\eeq
However, the correlations between $\ni$ and $\ne$ do not vanish under $\Ha_0$, because the solid interacts with water, which in turn interacts with the ions. 
As an approximation, we neglect these correlations, that is we neglect the correlation $\langle\ni \ne\rangle_0$. Physically, we assume we can separate the dielectric screening of the ions by water from the dielectric screening of the solid by water. 

Within this approximation, the force can be expressed as usual in terms of response functions, and higher-order diagrams can be systematically included \cite{Kavokine2022}. Thus, the friction coefficient becomes:
\beq\tilde \lambda_{\rm i}^{\rm FI}= \frac{k_{\rm B}T}{2 \pi^2} \int_0^{+\infty}q^3 \mathrm{d}q \int_0^{+\infty} \frac{\mathrm{d}\omega}{\omega^2} \frac{ \mathrm{Im}[\tilde g_{\rm e}^{\rm R}(q,\omega)] \, \mathrm{Im}[g_{\rm i}^{\rm R}(q,\omega)] }{|1-\tilde g_{\rm e}^{\rm R}(q,\omega)\,g_{\rm i}^{\rm R}(q,\omega)|^2},
\eeq
where the solid's response function is computed in \ti{presence} of water. 

At first order in the solid-water interaction, the solid's response function writes
\beq \tilde g_{\rm e}^{\rm R}(q,\omega)= g_{\rm e}^{\rm R}(q,\omega)(1-g_{\rm w}^{\rm R}(q,\omega))=\frac{g_{\rm e}^{\rm R}(q,\omega)}{\epsilon_{\rm w}(q,\omega)}\eeq
Therefore, the response of the solid is screened by the solvent. After taking the imaginary part, one gets 
\beq \im{\tilde g_{\rm e}^{\rm R}(q,\omega)}= \frac{\im{g_{\rm e}^{\rm R}(q,\omega)}}{\re{\epsilon_{\rm w}(q,\omega)}}+\re{g_{\rm e}^{\rm R}(q,\omega)}\im{g_{\rm w}^{\rm R}(q,\omega)}\eeq
The first term represents the friction on the solid through an interaction screened by the solvent. The second term represents the friction on water due to the electric potential induced by the solid. The latter does not correspond to an ion--solid interaction force and thus should be discarded when computing the ion--solid friction coefficient. Finally, 
\beq \boxed{\lambda_{\rm i}^{\rm FI}= \frac{k_{\rm B}T}{2 \pi^2} \int_0^{+\infty}q^3 \mathrm{d}q \int_0^{+\infty} \frac{\mathrm{d}\omega}{\omega^2\re{\epsilon_{\rm w}(q,\omega)}} \frac{ \mathrm{Im}[g_{\rm e}^{\rm R}(q,\omega)] \, \mathrm{Im}[g_{\rm i}^{\rm R}(q,\omega)] }{|1-\tilde g_{\rm e}^{\rm R}(q,\omega)\,g_{\rm i}^{\rm R}(q,\omega)|^2}}
\eeq
This approach is suitable for the theoretical investigation because we can estimate every term. 
However, approximations were made about the water screening that can be relaxed in the numerical treatment.

%%%%%%%%%%%%%%%%%%%%%%%%%%%%%%%%%%%%%%%%%%%%%%%%%%%%%%%%
\subsection{Perturbation theory with proper screening}

For this, we again carry out a perturbation theory, but with a different interaction Hamiltonian. Thus, we take:
\beq \Ha_0=\Ha_{\rm s}+\Ha_\ell \quad \tn{ and }\quad \Ha_{\rm int}=\Ha_{\rm w/s}+\Ha_{\rm i/s}\eeq
In this version, the solid does not interact with the electrolyte under $\Ha_0$. 

In linear response theory, the force on the ions becomes
\beq \langle\F_{\rm e\rightarrow i}\rangle(t)=\int \dd\x\dd\x'\dd t' \,\grad_\x\langle[\ni (\x)V(\x-\x')\ne(\x'),\Ha_{\rm int}(t')]\rangle_0\eeq
where the average is now taken over the Hamiltonian $\Ha_0$.
Thus, 
\beq \langle\F_{\rm e\rightarrow i}\rangle(t)= \int \dd\x\dd\x'\dd\bar\x\dd\bar\x'\dd t'\grad_\x V(\x-\x')V(\bar\x-\bar\x')\langle [\ni (\x)\ne(\x'),n_{\rm \ell}(\bar\x)\ne(\bar\x')]\rangle_0\eeq
where $n_{\rm \ell}=\nw+\ni $ is the total charge in the electrolyte. We observe that a correlation function between $\ni $ and $\ntot$ appears, which does not simplify into an ionic correlation function because the ion-water correlation does not vanish. 

Then, we need to define new correlation functions $\mc{C}_{\rm i/\ell}(\x,t) =\langle \ni(\x,t)\ntot(0,0)\rangle$ and their retarded, advanced and Keldysh versions. 
Note that $\mc{C}_{\rm i/\ell}^{\rm K}(\q,\w)$ still fulfills a fluctuation-dissipation theorem in its frame of reference:
\beq \mc{C}_{\rm i/\ell}^{\rm K}(\q,\w) = \frac{2k_{\rm B}T}{\hbar(\w-\q\cdot\v)}\left(\mc{C}_{\rm i/\ell}^{\rm R}(\q,\w)-\mc{C}_{\rm i/\ell}^{\rm A}(\q,\w)\right) \eeq
but may possess a real part since it is a cross-correlation. Indeed, $\mc{C}_{\rm i/\ell}^{\rm R}(\q,\w)$ and $\mc{C}_{\rm i/\ell}^{\rm A}(\q,\w)$ are not conjugate in general.

Since the densities $\ntot$ and $\ne$ are independent under $\Ha_0$, the force can be computed from the correlation functions. 
Namely, in Fourier space:
\beq \frac{\langle\F_{\rm e\rightarrow i}\rangle}{\mc{A}}=- \int\frac{\dd\q \dd\w}{(2\pi)^3}(i\q)  \frac{i\hbar}{2}\left(g_{\rm e}^{\rm K}(\q,\w)\mc{C}_{\rm i/\ell}^{\rm R}(\q,\w) +g_{\rm e}^{\rm A}(\q,\w)\mc{C}_{\rm i/\ell}^{\rm K}(\q,\w)\right)\eeq
where we now need to adapt the computation from \cite{Kavokine2022} to this correlation function. 
We always have that $\langle\F_{\rm e\rightarrow i}\rangle=-\langle\F_{\rm i\rightarrow e}\rangle$ and then 
\beq \frac{\langle\F_{\rm e\rightarrow i}\rangle}{\mc{A}}= 
\frac{i k_{\rm B}T \v}{8\pi^2}\int_0^\infty q^2\dd q \int\frac{\dd\w}{\w^2} \im{g_{\rm e}^{\rm R}(\q,\w)}\left(\mc{C}_{\rm i/\ell}^{\rm R}(\q,\w)-\mc{C}_{\rm i/\ell}^{\rm A}(\q,\w)\right)\eeq
Using that $[\mc{C}_{\rm i/\ell}^{\rm R,A}(\q,\w)]^*=\mc{C}_{\rm i/\ell}^{\rm R,A}(\q,-\w)$ we deduce
\beq \frac{\langle\F_{\rm e\rightarrow i}\rangle}{\mc{A}}= 
-\frac{k_{\rm B}T \v}{8\pi^2}\int_0^\infty q^2\dd q \int\frac{\dd\w}{\w^2} \im{g_{\rm e}^{\rm R}(\q,\w)}\left(\im{\mc{C}_{\rm i/\ell}^{\rm R}(\q,\w)}-\im{\mc{C}_{\rm i/\ell}^{\rm A}(\q,\w)}\right)\eeq
Here, we recognize the spectral function, which we define with an additional factor of $1/2$ for convenience (contrary to standard notation):
\beq \mc{A}_{\rm i/\ell} = \frac{\mc{C}_{\rm i/\ell}^{\rm R}-\mc{C}_{\rm i/\ell}^{\rm A}}{2} \eeq

Beyond linear response theory, higher order diagrams can be included systematically following \cite{Kavokine2022}, and the friction coefficient becomes:
\beq \boxed{\lambda_{\rm i}^{\rm FI}= \frac{k_{\rm B}T}{2 \pi^2} \int_0^{+\infty}q^3 \mathrm{d}q \int_0^{+\infty} \frac{\mathrm{d}\omega}{\omega^2} \frac{ \mathrm{Im}[g_{\rm e}^{\rm R}(q,\omega)] \im{\mc{A}_{\rm i/\ell}(\q,\w) }}{|1- g_{\rm e}^{\rm R}(q,\omega)\,g_{\rm \ell}^{\rm R}(q,\omega)|^2}}
\eeq
where the solid's response function is computed in \ti{absence} of water. 

In this method, there is no approximation about the interactions between the ions and the solid throughout the solvent. Moreover, the solid's response function is, as usual, computed in the absence of water, and there is no dielectric screening to add. The price to pay, however, is to compute the correlation between the ions and the solvent. Indeed, by the fluctuation-dissipation theorem, the cross-correlation function between the ions and the electrolyte is
\beq \mc{A}_{\rm i/\ell}(\q,\w)=\frac{e^2}{2\epsilon_0q}\frac{i\omega}{2k_{\rm B}T}S_{\rm i/\ell}(\q,\omega)\eeq
where 
\beq S_{\rm i/\ell}(\q,\omega)=\langle \ni (\q,\omega) \delta \ntot(q,\omega)^*\rangle_0=\langle  |\ni (q,\omega)|^2\rangle_0+\langle \ni (\q,\omega) \nw(\q,\omega)^*\rangle_0\eeq
The correlation function $\langle \ni  \ni \rangle_0$ is connected to the response function of ions, while the correlation function $\langle \ni  \nw\rangle_0$ is connected to the interactions between the ions and the solvent, which is hard to compute theoretically. In practice, the quantity entering the friction coefficient is:
\beq \im{\mc{A}_{\rm i/\ell}(\q,\w)}=\frac{e^2}{2\epsilon_0q}\frac{\omega}{2k_{\rm B}T}\re{S_{\rm i/\ell}(\q,\omega)}\eeq
and 
\beq \re{S_{\rm i/\ell}(\q,\omega)}=\langle  |\ni (q,\omega)|^2\rangle_0+\re{\langle \ni (\q,\omega) \nw(\q,\omega)^*\rangle_0}\eeq
Indeed, taking the real part of this cross-correlation provides the effective charge observed by the solid after screening by water, while the imaginary part would provide information about the dielectric friction between the ions and the solvent. 
In short, this properly accounts for the screening by water that we approximated by a factor $1/\re{\epsw}$ in the previous section. 
Since this cross-correlation is accessible numerically, we can use it to compute the ionic friction from perturbation theory directly and compare the approximate screening prescription with the full screening. 

%%%%%%%%%%%%%%%%%%%%%%%%%%%%%%%%%%%%%%%%%%%%%%%%%%%%%
\subsection{Effective screening in the bulk}

Here, we provide a proper derivation of the ratio between $\im{g_{\rm i}(q,\omega)}$ and $\im{\mc{A}_{\rm i/\ell}(q,\omega)}$ in the bulk.

We consider two mean-field subsystems occupying the same volume: ions
($\mathrm{i}$) and water ($\mathrm{w}$). Each subsystem has an intrinsic charge fluctuation
$n_\alpha^0$ and a polarization-induced contribution $n_\alpha^{\rm pol}$,
with $\alpha=\mathrm{i},\mathrm{w}$. The physical charge fluctuation is therefore
\begin{equation}
    n_\alpha = n_\alpha^0+n_\alpha^{\rm pol}.
\end{equation}

We assume that the intrinsic fluctuations of the two subsystems are
statistically independent,
\begin{equation}
    \langle n_{\rm i}^0 n_{\rm w}^0\rangle=0.
\end{equation}

\paragraph{Displacement field and polarization.}
A charge fluctuation $n_\alpha^0$ generates the Coulomb potential
\begin{equation}
    \phi_\alpha(\mathbf r,t)
    =
    \int d\mathbf r'\,
    V_C(\mathbf r-\mathbf r')\,n_\alpha^0(\mathbf r',t),
\end{equation}
and the corresponding displacement field
\begin{equation}
    \mathbf D_\alpha(\mathbf r,t)
    =
    -\epsilon_0\nabla\phi_\alpha(\mathbf r,t).
\end{equation}
By construction,
\begin{equation}
    \nabla\cdot\mathbf D_\alpha=n_\alpha^0.
\end{equation}

In a homogeneous bulk system we Fourier transform
$(\mathbf r,t)\rightarrow(\mathbf q,\omega)$. Using
$V_C(\mathbf q)=1/(\epsilon_0q^2)$,
\begin{equation}
    \mathbf D_\alpha(\mathbf q,\omega)
    =
    -i\frac{\mathbf q}{q^2}
    n_\alpha^0(\mathbf q,\omega),
    \qquad
    i\mathbf q\cdot\mathbf D_\alpha=n_\alpha^0.
\end{equation}

The displacement field generated by one subsystem polarizes the other.
We write
\begin{equation}
    \mathbf P_{\rm i}=\chi_{\rm i}\mathbf D_{\rm w},
    \qquad
    \mathbf P_{\rm w}=\chi_{\rm w}\mathbf D_{\rm i},
\end{equation}
where $\chi_{\rm i}(\mathbf q,\omega)$ and $\chi_{\rm w}(\mathbf q,\omega)$ are the
RPA-dressed susceptibilities. Since
$n_\alpha^{\rm pol}=-i\mathbf q\cdot\mathbf P_\alpha$, this gives
\begin{equation}
    n_{\rm i}^{\rm pol}=-\chi_{\rm i} n_{\rm w}^0,
    \qquad
    n_{\rm w}^{\rm pol}=-\chi_{\rm w} n_{\rm i}^0.
\end{equation}

For a susceptibility defined with respect to the displacement field, the
dielectric function is conveniently defined by
\begin{equation}
    \boxed{
    \epsilon_\alpha^{-1}(\mathbf q,\omega)
    =
    1-\chi_\alpha(\mathbf q,\omega).
    }
\end{equation}
This form already includes the RPA resummation contained in $\chi_\alpha$.

\paragraph{Ionic structure factor.}
The physical ionic fluctuation contains both its intrinsic fluctuation and
the response induced by water,
\begin{equation}
    n_{\rm i}=n_{\rm i}^0-\chi_{\rm i} n_{\rm w}^0.
\end{equation}
The ionic dynamical structure factor is defined from the full ionic
fluctuation,
\begin{equation}
    S_{\rm i}(\mathbf q,\omega)
    \equiv
    \left\langle
        n_{\rm i}(\mathbf q,\omega)
        n_{\rm i}(-\mathbf q,-\omega)
    \right\rangle.
\end{equation}
Using $\langle n_{\rm i}^0n_{\rm w}^0\rangle=0$,
\begin{equation}
    \boxed{
    S_{\rm i}
    =
    S_{\rm i}^0
    +
    |\chi_{\rm i}|^2 S_{\rm w}^0,
    }
\end{equation}
where
\begin{equation}
    S_\alpha^0
    =
    \langle n_\alpha^0 n_\alpha^0\rangle.
\end{equation}

\paragraph{Ionic--total cross-correlation.}
The total charge fluctuation is
\begin{equation}
    n_{\rm tot}
    =
    n_{\rm i}^0+n_{\rm i}^{\rm pol}+n_{\rm w}^0+n_{\rm w}^{\rm pol},
\end{equation}
thus
\begin{equation}
    n_{\rm tot}
    =
    (1-\chi_{\rm w})n_{\rm i}^0+(1-\chi_{\rm i})n_{\rm w}^0
    =
    \frac{n_{\rm i}^0}{\epsilon_{\rm w}}+ \frac{n_{\rm w}^0}{\epsilon_{\rm i}}.
\end{equation}

We define the ionic--total cross-spectrum by
\begin{equation}
    S_{\rm cross}(\mathbf q,\omega)
    =
    \left\langle
        n_{\rm i}(\mathbf q,\omega)
        n_{\rm tot}(-\mathbf q,-\omega)
    \right\rangle.
\end{equation}
Using the expression above for $n_{\rm tot}$ gives
\begin{equation}
    S_{\rm cross}
    =
    \frac{S_{\rm i}^0(\mathbf q,\omega)}
         {\epsilon_{\rm w}(\mathbf q,\omega)^*}
    -\frac{\chi_{\rm i}(\mathbf q,\omega)}{\epsilon_{\rm i}(\mathbf q,\omega)^*}S_{\rm w}^0(\mathbf q,\omega)
\end{equation}
that is,
\begin{equation}
    S_{\rm cross}
    =
    \frac{S_{\rm i}(\mathbf q,\omega)}{\epsilon_{\rm w}(\mathbf q,\omega)^*}
    -\left( \frac{|\chi_{\rm i}(\mathbf q,\omega)|^2}{\epsilon_{\rm w}(\mathbf q,\omega)^*} +  \frac{\chi_{\rm i}(\mathbf q,\omega)}{\epsilon_{\rm i}(\mathbf q,\omega)^*}\right)S_{\rm w}^0(\mathbf q,\omega)
\end{equation}

\paragraph{Interpretation.}

We denote by $g_{\rm i}(\mathbf q,\omega)$ the ionic response function and by
$A_{\rm i}(\mathbf q,\omega)$ the response associated with the ionic--total
cross-correlation. With the fluctuation--dissipation convention
\begin{equation}
    \operatorname{Im}g_{\rm i}
    =
    \frac{\omega}{2k_{\rm B}T}S_{\rm i},
    \qquad
    \operatorname{Im}A_{\rm i}
    =
    \frac{\omega}{2k_{\rm B}T}\operatorname{Re}S_{\rm cross},
\end{equation}
their ratio is
\begin{equation}
    \boxed{
    \frac{
        \operatorname{Im}A_{\rm i}(\mathbf q,\omega)
    }{
        \operatorname{Im}g_{\rm i}(\mathbf q,\omega)
    }
    =
    \operatorname{Re}
    \left[
        \frac{1}{\epsilon_{\rm w}(\mathbf q,\omega)} 
    \right]
    -\operatorname{Re}
    \left[
        \frac{|\chi_{\rm i}(\mathbf q,\omega)|^2}{\epsilon_{\rm w}(\mathbf q,\omega)^*} +  \frac{\chi_{\rm i}(\mathbf q,\omega)}{\epsilon_{\rm i}(\mathbf q,\omega)^*}
    \right] \frac{S_{\rm w}^0(\mathbf q,\omega)}{S_{\rm i}(\mathbf q,\omega)}
    }
\end{equation}
which we identify with $1/\epsilon_{\rm eff}(\mathbf q,\omega)$. The first term is the usual dielectric screening of water, while the second term is a higher-order correction in the ion--water interaction. This correction is encapsulated in the definition of the effective dielectric screening. 

Here, we performed this computation in the bulk, but the principle remains the same near the interface, with directions and distances to the interface consistently taken into account in the definitions of the correlation and response functions.

%%%%%%%%%%%%%%%%%%%%%%%%%%%%%%%%%%%%%%%%%%%%%%%%%%%%%
\subsection{Numerical evaluation of the effective screening}

%%%%%%%%%%%

In order to relate the ionic surface response function $g_{\rm i}(q,\omega)$ to the cross-correlation $\mc{A}_{\rm i/\ell}(q,\omega)$, we define in the main text the following effective screening:
\begin{equation}
    \epsilon_{\text{eff}}(q,\omega) = \frac{\im{g_{\rm i}(q,\omega)}}{\im{\mc{A}_{\rm i/\ell}(q,\omega)}}.
\end{equation}
To relate these quantities in practice, we assume that the permittivity follows a two-timescale relaxation model: 
\begin{equation}\label{eq:eff_perm_fit}
    \epsilon_{\rm eff}(q,\omega)-1 = \frac{\epsilon_{\rm eff}(q,0)-1}{2}\left[\frac{1}{1+\left(\frac{\omega}{\omega_1}\right)^2}+\frac{1}{1+\left(\frac{\omega}{\omega_2}\right)^2}\right],
\end{equation}
where $f_1 = \omega_1/2\pi = 0.03$ THz coincides with the Debye peak of water, $f_2 = \omega_2/2\pi = 1.4$ THz coincides with the higher-frequency relaxation mode, and $\epsilon_{\text{eff}}(q,0)$ is fitted by a decaying rational function:
\begin{equation}
    \epsilon_{\text{eff}}(q,0) = \frac{\epsilon_{\text{eff}}(0,0)+c\times q}{1+a\times q+b\times q^2},
\end{equation}
for $q$ in nm$^{-1}$. We obtained the following values for these coefficients:
\begin{table}[H]
    \centering
    \begin{tabular}{ c || c | c | c | c }
        $c_{\rm i}$ (M) & $\epsilon_{\rm eff}(0,0)$ & a & b & c\\
        \hline
        \hline
        1.6 & 16.31 & -0.351 & 0.0497 & -0.539 \\
        \hline
        3.2 & 13.71 & -0.288 & 0.0395 & -0.425 \\
        \hline
        6.4 & 8.873 & -0.149 & 0.0203 & -0.290
    \end{tabular}
    \caption{Fit parameters for the effective permittivity.}\label{tab:coeff_eff_screening}
\end{table}

We use the value for $c_{\rm i} = 6.4$~M in Fig. 3(e) and for $c_{\rm i} = 1.6$~M in Fig. 3(f) as an estimate of the dilute effective permittivity.

%%%%%%%%%%%%%%%%%%%%%%%%%%%%%%%%%%%%%%%%%%%%%%%%%%%%%
\section{Ionic modes}

%%%%%%%%%%%%%%%%%%%%%%%%%%%%%%%%%%%%%%%%%%%%%%%%%%%%%
\subsection{Poisson-Nernst-Planck formalism}

We consider a bulk electrolyte in 2D or 3D of salt density $c_{\rm i}$. 
We consider an external electric potential $\phi\ext$ applied to the electrolyte and first compute the bulk susceptibility $\chi_{\rm i}$, which gives the resulting ionic charge density distribution:
\beq\ni (\x,t)=\int\dd\x'\dd t'\, \chi_{\rm i}(\x,t,\x',t')\phi\ext(\x',t')\eeq

The charge-density dynamics is governed by the Nernst--Planck equation, which combines the continuity equations for both ionic species with Fick's law and the electric drift term:
\beq \partial_t \ni =\nabla\.\left[2Z\frac{D_{\rm i}}{k_{\rm B}T}c_{\rm i}\nabla \phi\ext+2Z\frac{D_{\rm i}}{k_{\rm B}T}c_{\rm i}\nabla \phi_{\rm ind}+D_{\rm i}\nabla\delta \ni \right]\eeq
Here, the electric potential experienced by the ions is the sum of the external electric potential and the induced field $\phi_{\rm ind}$, which describes screening by the other ions and is itself screened by the solvent. The field induced by the charge density is
\beq \phi_{\rm ind}(\x,t)=Z\int\dd\x'\, V_\epsilon(\x-\x')\ni (\x',t)\eeq
where $V_\epsilon$ is the screened Coulomb potential, which accounts for screening by the solvent.
In Fourier space, neglecting for simplicity the anisotropy and wavevector dependence of the solvent dielectric function $\epsw$, we obtain
\beq i\w\ni =2Zq^2\frac{D_{\rm i}}{k_{\rm B}T}c_{\rm i}\phi\ext+ \frac{ \omega_{\rm p}(q)}{\epsw (\w)}\ni  +q^2D_{\rm i}\ni \eeq
where 
\beq \omega_{\rm p}(q)= \frac{(Ze)^2}{\epsilon_0}\frac{D_{\rm i}}{k_{\rm B}T}\frac{c_{\rm i}^{\rm 2D}}{2}q \quad \tn{ in 2D,} \quad \tn{ and } \quad \omega_{\rm p}(q)= 2\frac{(Ze)^2}{\epsilon_0}\frac{D_{\rm i}}{k_{\rm B}T}c_{\rm i} \quad\tn{ in 3D}
\eeq
is the plasmon frequency. In the 2D case, the factor $1/2$ accounts for the two ionic layers and yields better agreement with our simulations.
Then, we deduce
\beq\ni (\q,\w)=-\frac{2c_{\rm i}}{k_{\rm B}T}\frac{ZD_{\rm i}q^2}{D_{\rm i}q^2+ \omega_{\rm p}(q)/\epsw (\w)-i\w}\phi\ext(\q,\w)\eeq
Finally, the susceptibility is: 
\beq \chi_{\rm i}(\q,\w)=-\frac{2c_{\rm i}}{k_{\rm B}T}\frac{ZD_{\rm i}q^2}{D_{\rm i}q^2+ \omega_{\rm p}(q)/\epsw (\w)-i\w}\eeq
The ionic response function describes the unscreened electric potential generated by the ionic charge density, thus:
\beq g_{\rm i}^{\rm PNP}(\q,\w)=- Zv_q\chi_{\rm i}(\q,\w)=  \frac{\omega_{\rm p}(q)}{D_{\rm i}q^2+\omega_{\rm p}(q)/\epsw (\w)-i\w}\eeq
For a 2D layer of ions, the plasmon dispersion is linear, whereas it is constant in a 3D bulk.
Here, the ionic distribution is made of 2 parallel quasi-2D layers, each with an effective 2D ionic concentration $\frac{c_{\rm i}^{\rm 2D}}{2} = c_{\rm i}\times \frac{h}{2}$ at a typical distance $d$ from the wall.
Thus, the surface response function is expected to take the form
\beq \boxed{g_{\rm i}^{\rm PNP}(\q,\w)= \frac{\omega_{\rm p}(q)}{D_{\rm i}q^2+\omega_{\rm p}(q)/\epsw (\w)-i\w}e^{-2qd}}\label{PNP_propagator}\eeq
with 
\beq  \omega_{\rm p}(q) \approx \frac{(Ze)^2}{\epsilon_0}\frac{D_{\rm i}}{k_{\rm B}T}\frac{c_{\rm i}}{2}q h = D_{\rm i}\kappa_0^2 \frac{q h}{4} \eeq
where $\kappa_0$ is the Debye wavevector computed with a vacuum dielectric background.

%%%%%%%%%%%%%%%%%%%%%%%%%%%%%%%%%%%%%%%%%%%%%%%%%%%%%
\subsection{Self part of the surface response function of the diffusive and plasmonic modes}\label{sec:si_numfit}

If we consider all the ionic species, we have:
\begin{equation}
    \im{g_{\rm i}(q,t)} \propto \left\langle\sum_{i,j}q_iq_je^{i\tb{q}\cdot(\tb{r}_i(t)-\tb{r}_j(0))}e^{-q(|z_i(t)-z_0|+|z_j(0)-z_0|)}\right\rangle,
\end{equation}
where the sum runs over all ions (coherent response). The response can be separated into two contributions, including the individual response (incoherent):
\begin{equation}
    \im{g_{\rm i}^{\rm s}(q,t)} \propto \left\langle\sum_{i}e^{i\tb{q}\cdot(\tb{r}_i(t)-\tb{r}_i(0))}e^{-q(|z_i(t)-z_0|+|z_i(0)-z_0|)}\right\rangle,
\end{equation}
with the remainder $g_{\rm i}^{\rm c}(q,t) = g_{\rm i}(q,t)-g_{\rm i}^{\rm s}(q,t)$.

Following the main text, we will model both the PNP contribution and the Debye cloud model contribution using the following function:
\begin{equation}
    g_{\rm i}^{\rm s}(q,\omega) = \frac{A(q)}{\omega_{\rm s}(q)-i\omega}+\frac{B(q)}{\omega_{\rm h}^2(q)-\omega^2-i\gamma(q)\omega},
\end{equation}
which represent a Debye relaxation and a harmonic relaxation, respectively.

\begin{figure}[H]
    \centering
    \includegraphics[width=\textwidth]{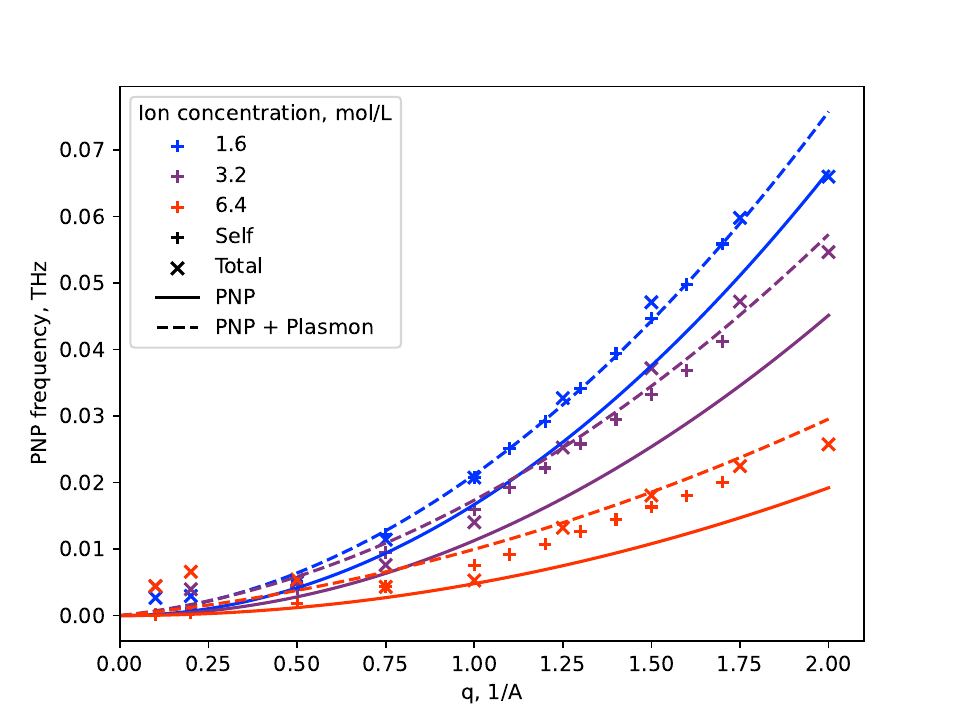}
    \caption{Frequency $\omega_{\rm s}$ of the PNP peak of the imaginary part of the surface response function. (Crosses) Using $\im{g_{\rm i}}$. (Plus signs) Using the self part only. (Solid line) Diffusive mode $\omega = D_{\rm i}q^2$. (Dashed line) PNP model $\omega = D_{\rm i}q^2+\omega_{\rm p}(q)$.}\label{fig:self_freq}
\end{figure}

We are particularly interested in the main frequency of the first peak, $\omega_{\rm s}(q)$. It should capture only individual processes, here diffusion, rather than collective effects. Indeed, the dispersion relation agrees very well with a diffusion process:
\begin{equation}
    \omega_{\rm s}(q) \propto q^2.
\end{equation}
However, as shown in Fig.~\ref{fig:self_freq}, the effective self-diffusion coefficient does not match the coefficient obtained from the mean-square displacement (solid curve), indicating that the self/collective splitting does not fully decouple the effects.

\begin{figure}[H]
    \centering
    \includegraphics[width=\textwidth]{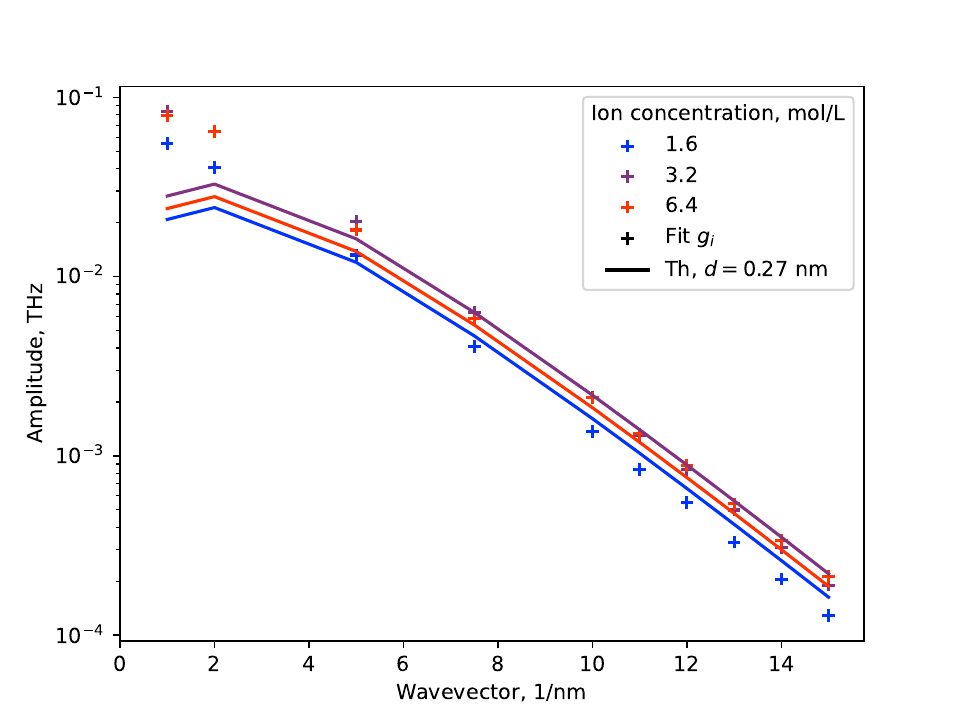}
    \caption{Amplitude of the PNP peak of the imaginary part of the surface response function. (Plus signs) Using the self part of $\im{g_{\rm i}}$. (Line) Amplitude $\omega_{\rm p}(q)e^{-2qd}$.}\label{fig:self_amp}
\end{figure}

The amplitude of the self part is shown in Fig.~\ref{fig:self_amp}. The agreement is good at high wavevectors, indicating that collective effects are negligible. At smaller wavevectors, the mismatch indicates strong collective effects.

%%%%%%%%%%%%%%%%%%%%%%%%%%%%%%%%%%%%%%%%%%%%%%%%%%%%%
\subsection{Debye cloud mode}

We consider an ion of charge $Q_{\rm i}=Ze$ and describe its screening by the other ions using a continuum model.
Thus, the electric potential $\phi_{\rm dc}$ created by the countercharges solves the Poisson--Boltzmann equation.
In the Debye--H\"uckel limit,
\beq \Delta \phi_{\rm dc} = \kappa^2\phi_{\rm dc}\eeq
Using spherical symmetry, and electroneutrality, we obtain a Yukawa potential:
\beq \phi_{\rm dc}(r) =  \frac{Q_{\rm i}}{4\pi \epsilon r}\left(e^{-\kappa r}-1\right) \eeq
From this potential, we can compute the electrostatic energy of the ion when it moves while the ionic cloud remains centered at $\x=0$.
For this, we need to take into account finite-size effects.
We then model the ion as a uniformly charged sphere of radius $a$. When its center is at position $r$, the energy is
\beq \mc{E}_{\rm dc}(r)=- \frac{Q_{\rm i}^2}{4\pi \epsilon}\frac{3}{4\pi a^3}\int_{|\x'|<a}\dd\x'\,\frac{e^{-\kappa |\x+\x'|}}{ |\x+\x'|} \eeq
Thus, 
\beq 
k_{\rm dc}=-\frac{1}{3}\Delta_\x \mc{E}_{\rm dc}(r=0)= \frac{Q_{\rm i}^2}{4\pi \epsilon \lambda_{\rm D}^2}\frac{1}{4\pi a^3}\int_{|\x'|<a}\dd\x'\, \frac{e^{-\kappa |\x'|}}{ |\x'|}
\eeq
where we used the fact that $\phi_{\rm dc}(r)$ is a solution of the Poisson--Boltzmann equation.
Then, 
\beq 
k_{\rm dc}= \frac{Q_{\rm i}^2}{4\pi \epsilon}\frac{1}{ a^3}\int_0^{\kappa a}\dd u\, u e^{-u} =  \frac{Q_{\rm i}^2}{4\pi \epsilon }\frac{1}{ a^3}\left[ 1-(1+\kappa a)e^{-\kappa a}\right]
\eeq
Assuming $\kappa a\ll 1$, we then obtain:
\beq 
k_{\rm dc}\approx   \frac{Q_{\rm i}^2}{8\pi \epsilon a \lambda_{\rm D}^2} = \frac{k_{\rm B}T \ell_{\rm B}Z^2}{2a\lambda_{\rm D}^2}
\eeq

The ionic cloud lags behind the ion and remains centered at $\x_{\rm dc}$.
Thus, in the presence of a small external force $\F$, the ion dynamics is given by
\beq m_{\rm i} \ddot \x = -k_{\rm dc} (\x-\x_{\rm dc})  +\F \eeq
Because the ionic cloud consists of mobile ions, it moves according to the ionic mobility under the force exerted by the central ion.
Thus
\beq \dot \x_{\rm dc} = \frac{D_{\rm i}}{k_{\rm B}T}k_{\rm dc} (\x-\x_{\rm dc})\eeq
Going to Fourier space, 
\beq\x_{\rm dc} =\frac{ \x}{1-i  \w/\gamma_{\rm dc}} \eeq
where $\gamma_{\rm dc}=k_{\rm dc}D_{\rm i}/k_{\rm B}T\approx \ell_{\rm B} D_{\rm i} / 2a\lambda_{\rm D}^2 \sim 10$ GHz.
At low frequencies, $\w\ll \gamma_{\rm dc}$, the force experienced by the central ion is
\beq -k_{\rm dc} (\x-\x_{\rm dc})\approx i\w\frac{k_{\rm dc}}{\gamma_{\rm dc}}\x =-\xi\dot\x \eeq
where $\xi = k_{\rm dc}/\gamma_{\rm dc}$ is a friction coefficient for ionic transport arising from ionic interactions \cite{Avni2022}.

Here, we are interested in high frequencies, so we do not use the quasistatic approximation.
The more general expression is therefore
\beq \x = \frac{1+i\gamma_{\rm dc}/\w}{k_{\rm dc}-m_{\rm i}\w^2-im_{\rm i}\gamma_{\rm dc}\w} \F\eeq
from which we deduce the resulting charge density response function:
\beq g_{\rm i}^{\rm dc}(\q,\w)=\frac{2\rho_{\rm i} Q_{\rm i}^2}{4\pi\epsilon_0 m_{\rm i}}\frac{1+i\gamma_{\rm dc}/\w}{\w_{\rm dc}^2-\w^2-i  \gamma_{\rm dc} \w}e^{-2qd} \eeq
where $\w_{\rm dc}=\sqrt{k_{\rm dc}/m_{\rm i}}$. Close to the peak, $\w\approx \w_{\rm dc}\gg \gamma_{\rm dc}$, we can approximate the response function as
\beq \boxed{g_{\rm i}^{\rm dc}(\q,\w)\approx \frac{2\rho_{\rm i} Q_{\rm i}^2}{4\pi\epsilon_0 m_{\rm i}}\frac{1}{\w_{\rm dc}^2-\w^2-i  \gamma_{\rm dc} \w}e^{-2qd} }\eeq
which is a harmonic mode.
Its imaginary part reads:
\beq \im{g_{\rm i}^{\rm dc}(\q,\w)}\approx \frac{2\rho_{\rm i} Q_{\rm i}^2}{4\pi\epsilon_0 m_{\rm i}}\frac{\gamma_{\rm dc} \w}{[\w_{\rm dc}^2-\w^2]^2 +  (\gamma_{\rm dc} \w)^2}e^{-2qd} \eeq
which is a harmonic peak centered around
\beq \w \approx \w_{\rm dc} \approx \sqrt{\frac{k_{\rm B}T \ell_{\rm B}Z^2}{2a\lambda_{\rm D}^2 m_{\rm i}}} \eeq
 of amplitude 
\beq  \im{g_{\rm i}^{\rm dc}(\q,\w_{\rm dc})}\approx \frac{2\rho_{\rm i} Q_{\rm i}^2 \lambda_{\rm D}^3}{4\pi\epsilon_0 D_{\rm i}}\sqrt{\frac{8a^3 }{k_{\rm B}T \ell_{\rm B}^3Z^6 m_{\rm i}}}e^{-2qd} \eeq

Taking $a$ to be a hydrated radius, we set $a\approx 5$~\AA. For $c_{\rm i}=3$~M, we then find $\w_{\rm dc}\approx 1$~THz.

%%%%%%%%%%%%%%%%%%%%%%%%%%%%%%%%%%%%%%%%%%%%%%%%%%%%%%%%
\subsection{Hydration mode}

Similarly, the ion induces a solvation shell of water molecules that screens its charge.
The electric potential created by the polarization of water is
\beq \phi_{\rm h}(r)=-\frac{Q_{\rm i}}{4\pi\epsilon_0 r}\left(1-\frac{1}{\epsw(0) }\right) \eeq
by definition of the relative dielectric constant.
From this potential, we can compute the electrostatic energy of the ion when it moves while the solvation shell remains centered at $\x=0$.
For this, we need to take into account finite-size effects. 
We then model the ion as a uniformly charged sphere of radius $a$. When its center is at position $r$, the energy is
\beq \mc{E}_{\rm h}(r)=- \frac{Q_{\rm i}^2}{4\pi \epsilon_0}\left(1-\frac{1}{\epsw(0) }\right) \frac{3}{4\pi a^3}\int_{|\x'|<a}\dd\x'\,\frac{1}{ |\x+\x'|} \eeq
Thus, 
\beq 
k_{\rm h}=-\frac{1}{3}\Delta_\x \mc{E}_{\rm h}(r=0)=  \frac{Q_{\rm i}^2}{ \epsilon_0}\left(1-\frac{1}{\epsw(0) }\right)\frac{1}{4\pi a^3}\int_{|\x'|<a}\dd\x'\, \delta(\x') =  \frac{Q_{\rm i}^2}{4\pi \epsilon_0 a^3}\left(1-\frac{1}{\epsw(0) }\right)
\eeq
using that the Coulomb potential is the Green's function of the Poisson equation.
Since $\epsw(0) \gg1$, we can approximate:
\beq 
k_{\rm h} \approx  \frac{Q_{\rm i}^2}{4\pi \epsilon_0 a^3} = \frac{k_{\rm B}T\ell_{\rm B} \epsw(0) Z^2}{a^3}
\eeq

The solvation shell lags behind the ion and remains centered at $\x_{\rm h}$.
Thus, in the presence of a small external force $\F$, the ion dynamics is given by
\beq m_{\rm i} \ddot \x = -k_{\rm h} (\x-\x_{\rm h})  +\F \eeq
The solvation shell reorients through the Debye mode, so
\beq \dot \x_{\rm h} = \w_D(\x-\x_{\rm h})\eeq
Going to Fourier space, 
\beq\x_{\rm h} =\frac{ \x}{1-i  \w/\w_D} \eeq
At low frequencies, $\w\ll \w_D$, the force is
\beq -k_{\rm h} (\x-\x_{\rm h})\approx i\w\frac{k_{\rm h}}{\w_D}\x =-\xi\dot\x \eeq
where $\xi = k_{\rm h}/\w_D$ is the dielectric friction \cite{Bagchi1991, Bagchi1998, Balos2020}.

Here, we are interested in high frequencies, so we do not use the quasistatic approximation.
The more general expression is therefore
\beq \x = \frac{1+i\w_D/\w}{k_{\rm h}-m_{\rm i}\w^2-im_{\rm i}\w_D\w} \F\eeq
from which we deduce the resulting charge density response function:
\beq g_{\rm i}^{\rm h}(\q,\w)=\frac{2\rho_{\rm i} Q_{\rm i}^2}{4\pi\epsilon_0 m_{\rm i}}\frac{1+i\w_D/\w}{\w_{\rm h}^2-\w^2-i  \w_D \w}e^{-2qd} \eeq
where $\w_{\rm h}=\sqrt{k_{\rm h}/m_{\rm i}}$. Close to the peak, $\w\approx \w_{\rm h}\gg \w_D$, we can approximate the response function as
\beq \boxed{ g_{\rm i}^{\rm h}(\q,\w)\approx \frac{2\rho_{\rm i} Q_{\rm i}^2}{4\pi\epsilon_0 m_{\rm i}}\frac{1}{\w_{\rm h}^2-\w^2-i  \w_D \w}e^{-2qd}} \eeq
which is a harmonic mode.
Its imaginary part reads:
\beq \im{g_{\rm i}^{\rm h}(\q,\w)}\approx \frac{2\rho_{\rm i} Q_{\rm i}^2}{4\pi\epsilon_0 m_{\rm i}}\frac{\w_D \w}{[\w_{\rm h}^2-\w^2]^2 +  (\w_D \w)^2}e^{-2qd} \eeq
which is a harmonic peak centered around
\beq \w \approx \w_{\rm h} \approx \sqrt{ \frac{k_{\rm B}T\ell_{\rm B}Z^2 \epsw(0) }{a^3 m_{\rm i}}} \eeq
 of amplitude 
\beq  \im{g_{\rm i}^{\rm h}(\q,\w_{\rm h})}\approx \frac{2\rho_{\rm i} Q_{\rm i}^2}{4\pi\epsilon_0 \w_D}\sqrt{ \frac{a^3 }{k_{\rm B}T\ell_{\rm B}Z^2 \epsw(0)  m_{\rm i}}}e^{-2qd} \eeq

Taking $a$ to be the ion radius, we set $a\approx 2$~\AA. We then find $\w_{\rm h}\approx 20$~THz.

%%%%%%%%%%%%%%%%%%%%%%%%%%%%%%%%%%%%%%%%%%%%%%%%%%%%%
\subsection{Numerical fit of the harmonic modes}
\begin{figure}[H]
    \centering
    \includegraphics[width=\textwidth]{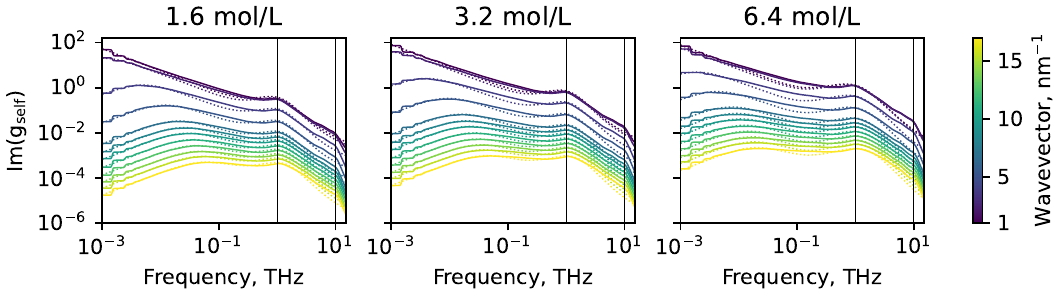}
    \caption{Imaginary part of the self surface response function. Dotted line: fit using PNP and a harmonic mode around 1 THz.}
    \label{fig:si_self}
\end{figure}

\begin{figure}[H]
    \centering
    \includegraphics[width=\textwidth]{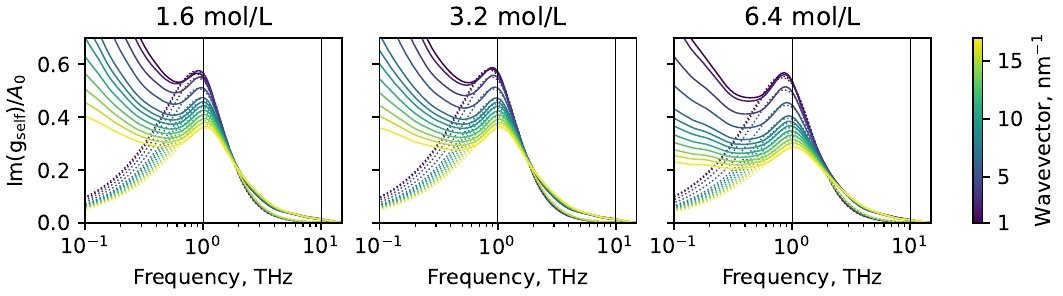}
    \caption{Imaginary part of the self surface response function. Dotted line: fit of a harmonic mode around 1 THz. The y-axis is normalized by the amplitude of the harmonic mode.}
    \label{fig:si_debye}
\end{figure}

We can use the preceding formula to fit the remainder of the spectrum. The Debye-cloud mode is reproduced very accurately by a harmonic mode:
\begin{equation}
    g_{\rm i}(\q,\w) = \frac{A(q)}{B(q)^2-\omega^2-i\gamma(q)\omega}.
\end{equation}
The results of a fit using a PNP mode and the Debye-cloud mode are shown in Fig.~\ref{fig:si_self}; a close-up is shown in Fig.~\ref{fig:si_debye}.

\begin{figure}[H]
    \centering
    \includegraphics[width=\textwidth]{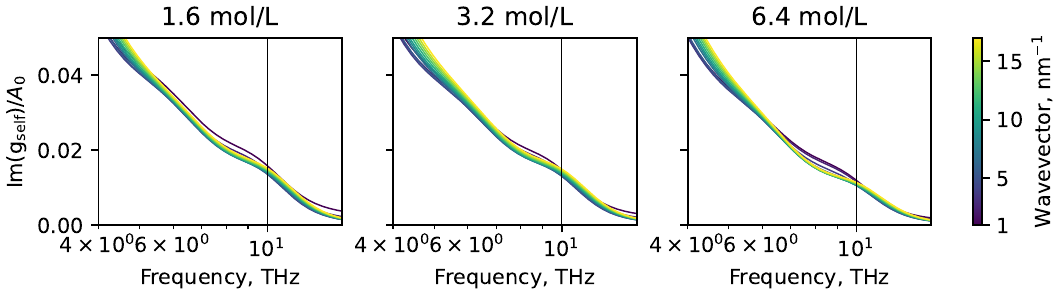}
    \caption{Imaginary part of the self surface response function. The y-axis is normalized by the amplitude of the harmonic mode at 1 THz.}
    \label{fig:si_hydration}
\end{figure}

The hydration mode is more difficult to fit with a simple harmonic peak. A close-up of this peak, normalized by the amplitude of the Debye-cloud mode for visibility, is shown in Fig.~\ref{fig:si_hydration}.

%%%%%%%%%%%%%%%%%%%%%%%%%%%%%%%%%%%%%%%%%%%%%%%%%%%%%

\section{Friction estimated from the fluctuation-induced formula}

%%%%%%%%%%%%%%%%%%%%%%%%%%%%%%%%%%%%%%%%%%%%%%%%%%%%%
\subsection{General usage}
For a subset of species $\alpha$, the fluctuation-induced friction formula is
\begin{equation}
    \lambda_{\alpha}^{\rm FI} = \frac{k_BT}{2\pi^2}\int_0^{\infty}\frac{q^3}{\omega^2}\frac{\im{g_{\rm dr}(q,\omega)}\im{\mathcal{A}_{\alpha/\ell}(q,\omega)}}{|1-g_{\rm dr}(q,\omega)g_{\ell}(q,\omega)|^2}\dd q\dd\omega.
\end{equation}
For the Drude spectrum, we use the theoretical formula derived in Sec.~\ref{sec:drude} with the correction factor of 2.

When computing the total, water, or ionic friction from the numerical spectrum, $\im{\mathcal{A}_{\alpha/\ell}(q,\omega)}$ is computed directly from simulations using the fluctuation--dissipation theorem. In the denominator, the imaginary part of $g_{\ell}$ can also be obtained from simulations, while the real part follows from the Kramers--Kronig relation. Because the imaginary part is an odd function of $\omega$,
\begin{equation}
    \re{g(q,\omega)} = \frac{2}{\pi}\mathcal{P}\int_0^{\infty}\frac{\omega'\im{g}(q,\omega')}{\omega'^2-\omega^2}\dd\omega'.
\end{equation}

When computing the friction from the theoretical model (PNP for the ions), we estimate $\im{\mathcal{A}_{{\rm i}/\ell}(q,\omega)}$ using the fitted effective permittivity. However, there is no straightforward way to obtain $g_{\ell}$. We therefore use only the first-order friction formula:
\begin{equation}
    \lambda_{\rm PNP}^{\rm FI} \sim \frac{k_BT}{2\pi^2}\int_0^{\infty}\frac{q^3}{\omega^2}\im{g_{\rm dr}(q,\omega)}\frac{\im{g_{\rm i}^{\rm PNP}(q,\omega)}}{\varepsilon_{\rm eff}(q, \omega)}\dd q\dd\omega.
\end{equation}
This generally slightly overestimates the friction [see Fig.~4(a) of the main text], but the approximation appears reasonable for $\omega\sim 0.1$~THz.

%%%%%%%%%%%%%%%%%%%%%%%%%%%%%%%%%%%%%%%%%%%%%%%%%%%%%

\subsection{Comparison of electrolyte friction obtained from nonequilibrium simulations and fluctuation-induced friction}

\begin{figure}[H]
    \centering
    \includegraphics[width=0.8\textwidth]{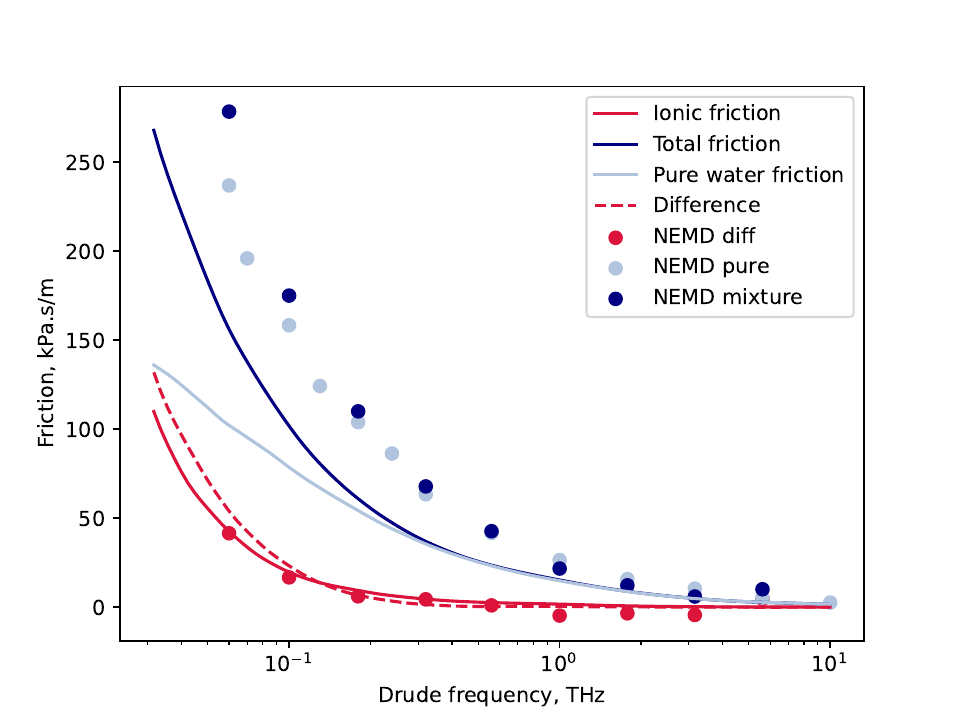}
    \caption{Comparison of the friction obtained from nonequilibrium simulations (dots) and the fluctuation-induced friction formalism (solid curves) for pure water (light blue) and a 6.4~M electrolyte (blue). The difference between these two curves is shown in red (dots and dashed curve). The solid red curve is obtained by applying the fluctuation-induced friction formalism to the ionic spectrum.}
    \label{fig:si_nemd_vs_qf}
\end{figure}

Nonequilibrium molecular dynamics simulations and the fluctuation-induced friction formalism provide two different methods for computing electrolyte friction. Figure~\ref{fig:si_nemd_vs_qf} compares the results for pure water and a 6.4~M electrolyte.

The two predictions differ substantially for the total electrolyte friction. Because the pure-water friction obtained from the fluctuation-induced friction formalism (light-blue solid curve) agrees well with the pure-water friction coefficient computed near an interface using Green--Kubo \cite{Bui2023} or nonequilibrium \cite{Coquinot2025} methods, we believe that our nonequilibrium simulations overestimate the friction coefficient. This discrepancy likely arises from a confinement effect that we do not fully understand. However, the two predictions agree very well for the ionic friction, so this issue does not affect the conclusions of this paper.

%%%%%%%%%%%%%%%%%%%%%%%%%%%%%%%%%%%%%%%%%%%%%%%%%%%%%

\subsection{Water friction for various ion concentrations}
In nonequilibrium molecular dynamics simulations, there is no simple way to separate the ionic and water contributions to friction, so we assume that
\begin{equation}
    \lambda_{\rm w}(c) \sim \lambda_{\rm w}(c=0).
\end{equation}

\begin{figure}[H]
    \centering
    \includegraphics[width=0.8\textwidth]{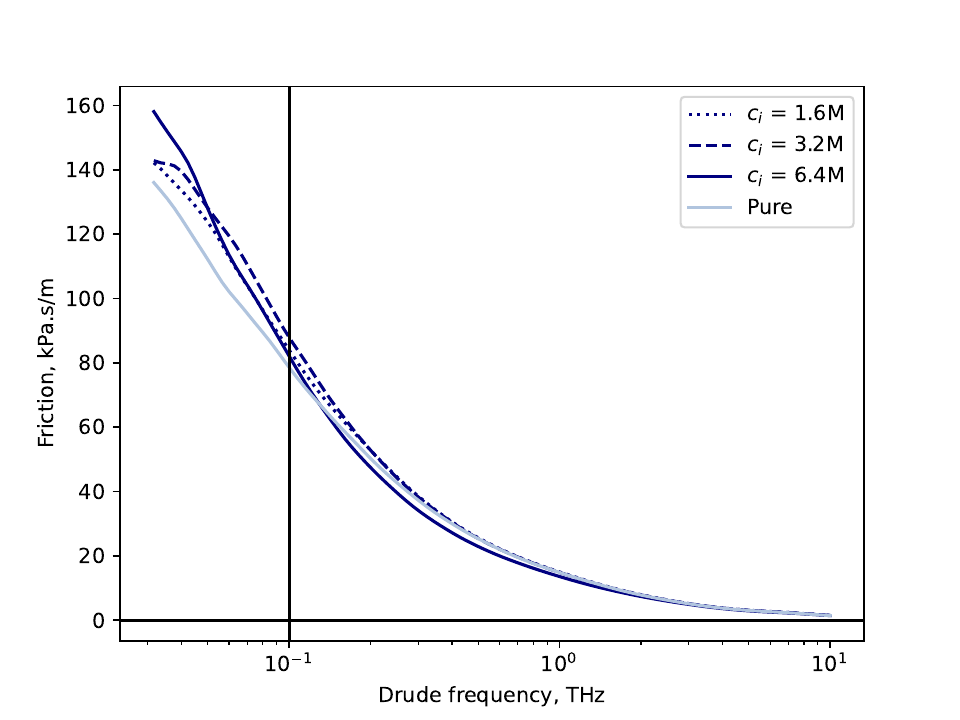}
    \caption{Fluctuation-induced friction of pure water (light blue) and water at various electrolyte concentrations (blue curves).}
    \label{fig:si_water_friction}
\end{figure}

We test this assumption using the fluctuation-induced friction formalism. Figure~\ref{fig:si_water_friction} shows the fluctuation-induced friction for pure water and water at various electrolyte concentrations.

Overall, the difference is small, which justifies our approximation. In particular, at the frequency $f\sim 0.1$~THz considered in the applications, using pure water as a reference overestimates the ionic friction obtained from NEMD by approximately 10~kPa$\cdot$s/m.

%%%%%%%%%%%%%%%%%%%%%%%%%%%%%%%%%%%%%%%%%%%%%%%%%%%%%

\subsection{Effect of confinement on the friction}

As described in Ref.~\cite{Coquinot2023b}, the quantum-friction formula should be modified under confinement because the water is sufficiently confined to be dragged by both walls. We computed the symmetric and antisymmetric water responses in simulations and deduced the corresponding friction coefficients. However, we observed no difference from the standard quantum-friction calculation.

%%%%%%%%%%%%%%%%%%%%%%%%%%%%%%%%%%%%%%%%%%%%%%%%%%%%%

\subsection{First-order and full-order fluctuation-induced formulas for NaCl and LiCl friction}

In Fig.~4(b) of the main text, we plot only the first-order fluctuation-induced friction formula:
\begin{equation}
    \lambda^{\rm FI, 1st} = \frac{k_BT}{2\pi^2}\int_0^{\infty}\frac{q^3\,\dd q\,\dd\omega}{\omega^2}\im{g_{\rm s}(q,\omega)}\im{g_{\ell}(q, \omega)}.
\end{equation}
\begin{figure}[H]
    \centering
    \includegraphics[width=0.8\textwidth]{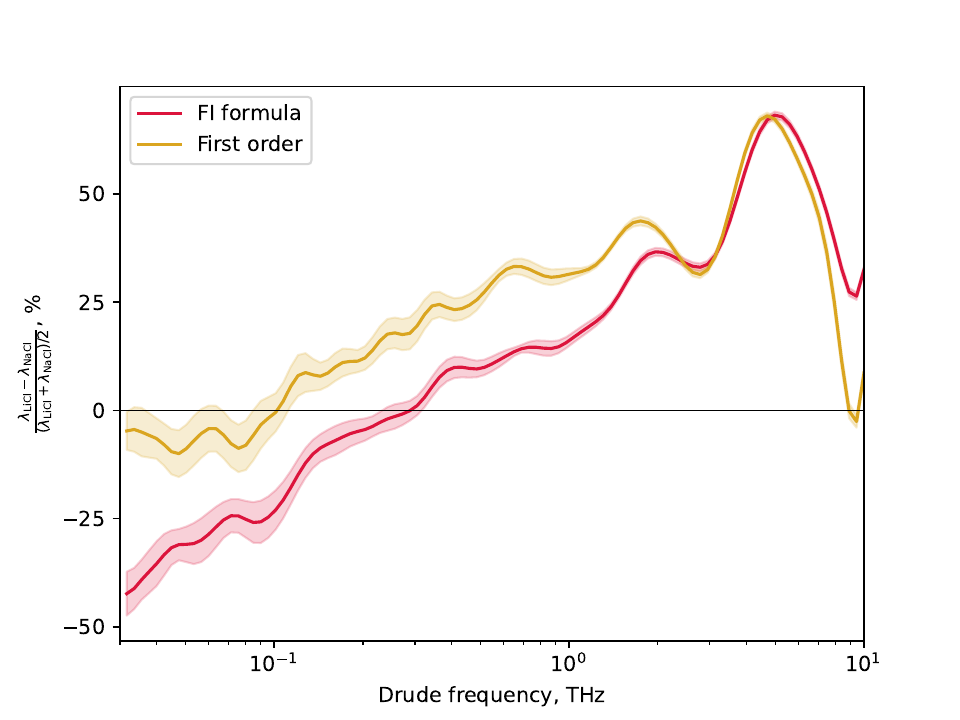}
    \caption{Relative difference in the friction coefficient between NaCl and LiCl at 3.2~M, computed using the first-order (yellow) and full-order (red) formulas.}
    \label{fig:si_order_diff_nali}
\end{figure}
A comparison of the results obtained using the first-order and full-order formulas is shown in Fig.~\ref{fig:si_order_diff_nali}. For $\omega_D\gtrsim 1$~THz, the difference is small, supporting the discussion in the main text. At lower frequencies, however, the full-order formula predicts greater friction for NaCl than for LiCl, whereas the first-order formula predicts no difference. This mismatch, together with the behavior of PNP at low frequencies, where the response becomes nonspecific to the ion, indicates that the additional friction is caused by a change in $g_{\ell}$ in the denominator of the full-order formula. Thus, the measured friction difference is not a difference in ionic friction but a difference in water friction between ion species. This effect may be related to changes in water-transport properties in the continuum limit at high ion concentrations, as studied in Ref.~\cite{Robin2024}. In the main text, we investigate the use of ionic friction to develop ion-separation technologies. For such applications, the relevant quantity is the friction of one ion relative to that of another, which is unaffected by a global change in water transport. We therefore restrict ourselves to the first-order formula.

%%%%%%%%%%%%%%%%%%%%%%%%%%%%%%%%%%%%%%%%%%%%%%%%%%%%%

\section{Physisorbed ions and permeability engineering}

%%%%%%%%%%%%%%%%%%%%%%%%%%%%%%%%%%%%%%%%%%%%%%%%%%%%%
\subsection{Comparison of NaCl and LiCl}
\begin{figure}[H]
    \centering
    \includegraphics[width=0.8\textwidth]{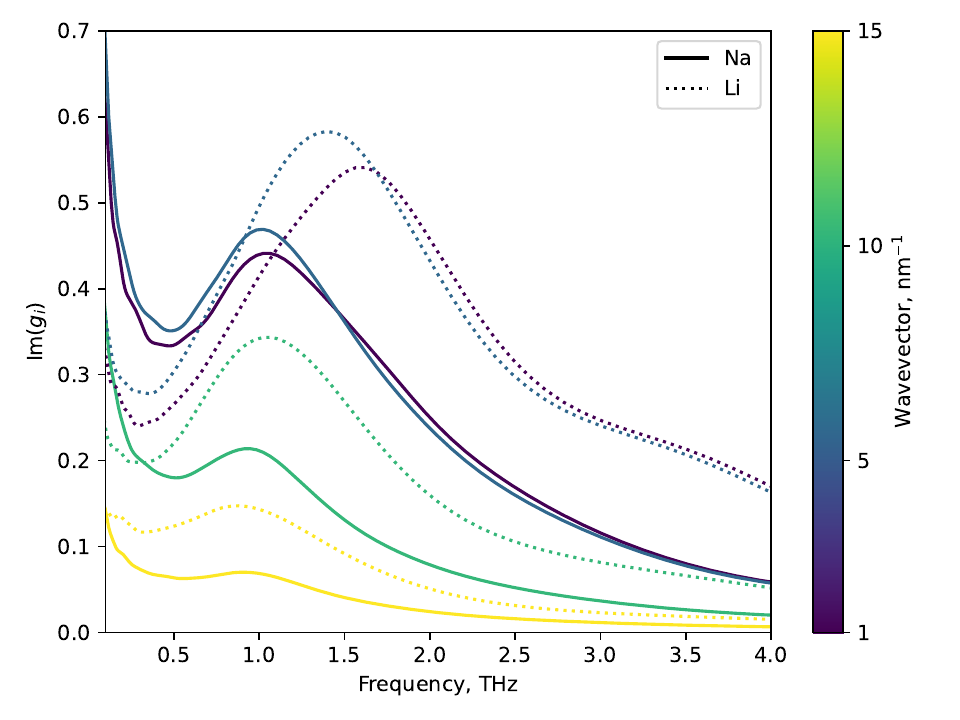}
    \caption{Imaginary part of the surface response function of NaCl (solid lines) and LiCl (dotted lines).}
    \label{fig:si_nali}
\end{figure}

A comparison of the Debye-cloud peaks for LiCl and NaCl is shown in Fig.~\ref{fig:si_nali}. The LiCl peak is shifted to the left, as expected because lithium has a lower mass and a smaller radius than sodium. From these spectra, we can compute the predicted friction difference plotted in the main text.

%%%%%%%%%%%%%%%%%%%%%%%%%%%%%%%%%%%%%%%%%%%%%%%%%%%%%
\subsection{Surface charge regulation}

Let us consider hydroxide ions subject to an adsorption potential $U_{\text{abs}}$.
The surface charge is fixed by the equilibrium of the chemical potential between adsorbed and free ions~\cite{Secchi2016b}:
\begin{equation}
    \mu = k_{\rm B}T\log\left|\frac{\Sigma}{e}\lambda^2\right|- e\phi_s- U_{\text{abs}},
\end{equation}
Here, $\lambda$ is a microscopic lengthscale, $\phi_s$ is the electric potential at the interface, and $\mu$ is fixed by the reservoir.
Therefore,
\begin{equation}\label{eq:s_of_p}
    \Sigma = -e \lambda c_010^{\rm pH} e^{(e\phi_s+ U_{\text{abs}})/k_{\rm B}T},
\end{equation}
where $c_0=1$ M is the reference concentration.
The electric potential at the interface, $\phi_s$, is obtained by solving the Poisson--Boltzmann equation in the nanochannel. Defining the reduced potential $\widetilde{\phi}=e\phi/k_{\rm B}T$, we have
\begin{equation}\label{eq:pb}
    \frac{\partial^2\widetilde{\phi}}{\partial z^2} = \kappa^2\sinh(\widetilde{\phi}),
\end{equation}
with the boundary conditions $\partial_z\widetilde{\phi}(\pm h/2)=\pm e\Sigma/\epsilon_0\epsilon_{\rm w} k_{\rm B}T$ controlled by the surface charge.
Eqs.~\eqref{eq:s_of_p}-\eqref{eq:pb} must be solved self-consistently.

%%%%%%%%%%%%%%%%%%%%%%%%%%%%%%%%%%%%%%%%%%%%%%%%%%%%%
\subsection{Debye--Hückel regime}
\begin{figure}[H]
    \centering
    \includegraphics[width=0.8\textwidth]{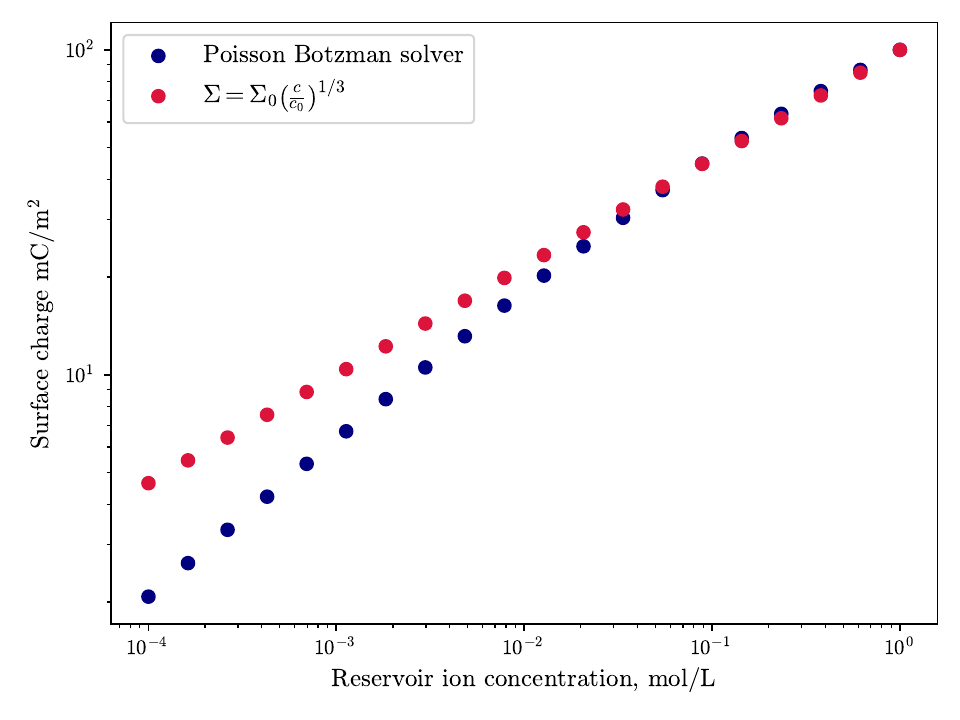}
    \caption{Surface charge as a function of the reservoir ion concentration, computed by solving the Poisson--Boltzmann equation (blue) or using the Debye--H\"uckel scaling (red).}
    \label{fig:si_sigma}
\end{figure}

In the Debye--H\"uckel regime, the surface charge scales as $\Sigma\propto c_{\rm i}^{1/3}$~\cite{Secchi2016b}. We choose $U_{\rm abs}$ and the pH such that $\Sigma(c = c_0 = 1~\textrm{M}) = \Sigma_0 = 100$~mC/m$^2$ and compare the result with the scaling
\begin{equation}
    \Sigma(c) = \Sigma_0\left(\frac{c}{c_0}\right)^{1/3}.
\end{equation}
The result is shown in Fig.~\ref{fig:si_sigma}. The excellent agreement supports the use of the Debye--H\"uckel approximation in the main text.

%%%%%%%%%%%%%%%%%%%%%%%%%%%%%%%%%%%%%%%%%%%%%%%%%%%%%
\subsection{Numerical procedure used to obtain Fig.~5}

%It is not straightforward in general to deduce the concentration of ions close to the surface from the surface charge. Here, we assume that only one type of ions can absorb to the surface (here for our absorption parameters hydroxyle ions). 
We use a frequency of 0.1~THz, for which we have already estimated the friction as a function of the concentration in the slit. At such a low frequency, all ion-specific effects should disappear. More specifically, we assume PNP and plasmon modes with a diffusion coefficient obtained from simulations of system I (see Table~\ref{tab:syst} and Sec.~\ref{sec:si_diff}). The resulting diffusion coefficients are shown in Table~\ref{tab:diff} and are fitted and extrapolated using the linear law $D(c) = D_0-\alpha c$.

\begin{table}[H]
    \centering
    \begin{tabular}{ c || c | c | c }
        Concentration (M) & 1.6 & 3.2 & 6.4 \\
        \hline
        Diffusion coefficient ($10^{-9}$ m$^2$/s) & 1.379 & 1.1 & 0.516\\
    \end{tabular}
    \caption{Diffusion coefficient obtained from system III.}\label{tab:diff}
\end{table}

We then assume that the effective permittivity obtained for $c_{\rm i} = 1.6$~mol/L remains valid at lower concentrations and use it to deduce the screened surface response function. Finally, we compute the friction using the fluctuation-induced friction formula. The total distance $d_{\rm tot}$ between the ions and the wall is defined as follows. For the force-field parameters used here, the distance $d_{\rm dr}$ between the wall and the liquid--solid interface is 0.13~nm \cite{Kavokine2022}. In our system, the distance $d_{\rm ion}$ between the ion and the interface is 0.27~nm (see Supplemental Material, Sec.~3.1), giving a total distance of 0.4~nm. We also estimate the friction for ions 0.05~nm closer to the wall, which is more realistic for physisorbed ions.

Finally, the slip length is computed using:
\begin{equation}
    b = \frac{\eta}{\lambda_{\rm i}+\lambda_{\rm w}},
\end{equation}
taking $\eta = 0.8509\times 10^{-3}$~Pa$\cdot$s \cite{Aleksandrov2008} and $\lambda_{\rm w} = 78.0$~kPa$\cdot$s/m at 0.1~THz (for pure water, see Supplemental Material, Sec.~6.2).

\bibliographystyle{plain}
\bibliography{bibfile}